\documentclass[a4paper,11pt]{article}
\pdfoutput=1
\usepackage{jheppub}
\usepackage[utf8x]{inputenc}
\usepackage{amssymb}
\usepackage{bm}
\usepackage{multirow}
\usepackage{subfigure}
\usepackage{slashed}

\newcommand{\be}{\begin{equation}}
\newcommand{\ee}{\end{equation}}
\newcommand{\ba}{\begin{eqnarray}}
\newcommand{\ea}{\end{eqnarray}}
\newcommand{\bal}{\begin{align}}
\newcommand{\eal}{\end{align}}
\newcommand{\bfig}{\begin{figure}}
\newcommand{\efig}{\end{figure}}

\newcommand{\rmdx}[1]{\mbox{d} #1 \,} 
\newcommand{\rmDx}[1]{\mbox{$\mathcal{D}$} #1 \,} 
\DeclareMathOperator{\argmax}{arg max}
\makeatletter
\let\Im\@undefined
\makeatother
\DeclareMathOperator{\Im}{Im}

\newcommand{\MSbar}{\ensuremath{\overline{\text{MS}}}}
\newcommand{\mbMSbar}{\overline{m}_b(\overline{m}_b)}
\newcommand{\GeV}{\mbox{GeV}}
\newcommand{\MeV}{\mbox{MeV}}

\newcommand{\refeq}[1]{eq.~(\ref{eq:#1})}
\newcommand{\reffig}[1]{figure~\ref{fig:#1}}
\newcommand{\reftab}[1]{table~\ref{tab:#1}}
\newcommand{\refsec}[1]{section~\ref{sec:#1}}

\title{Extrapolation and unitarity bounds\\ for the $B\to \pi$ form factor}
\author{I. Sentitemsu Imsong,}
\emailAdd{imsong@physik.uni-siegen.de}
\author{Alexander Khodjamirian,}
\emailAdd{khodjamirian@physik.uni-siegen.de}
\author{Thomas Mannel,}
\emailAdd{mannel@physik.uni-siegend.de}
\author{Danny van Dyk}
\emailAdd{vandyk@physik.uni-siegen.de}
\affiliation{Theoretische Physik 1, Naturwissenschaftlich-Technische Fakult\"at, Universit\"at Siegen, D-57068 Siegen, Germany}

\abstract{%
We address the problem of extrapolating the vector form factor $f_{B\pi}^+$,
which is relevant to $B\to \pi\ell \nu_\ell$
decays, from the region of small to the region of large momentum transfer.
As input, we use the QCD light-cone sum rule at small momentum transfer.
We carry out a comprehensive Bayesian
uncertainty analysis and obtain correlated uncertainties for
the normalization and shape parameters of the form factor.
The $z$-series parametrization for $f_{B\pi}^+$
is employed
to extrapolate our results to large momentum transfer, and to compare
with the lattice QCD results.
To test the validity of our extrapolation
we  use the upper and lower bounds from the unitarity
and positivity of the two-point correlator of heavy-light quark currents.
This correlator is updated by including the NNLO perturbative term and the
NLO correction to the quark condensate contribution.
We demonstrate that an additional input including the form factor, its
first and second derivative
calculated  at one value of momentum transfer
from the light-cone sum rules, considerably improves the bounds. This only holds when
the correlations between the form factor parameters are taken into account.
We further combine our results with the latest
experimental measurements of $B\to \pi \ell \nu_\ell$
by the BaBar and Belle collaborations, and obtain $|V_{ub}|= (3.32^{+0.26}_{-0.22}) \cdot 10^{-3}$
from a Bayesian analysis.
}


\preprint{SI-HEP-2014-21, QFET-2014-14, EOS-2014-02}

\begin{document}

\maketitle


\section{Introduction \label{sec:Intro}}

Hadronic form factors that are relevant to exclusive semileptonic $B$ decays remain in the center of attention.
The $B\to \pi$ vector form factor $f^+_{B\pi}(q^2)$
in the semileptonic region of the momentum transfer,  $0<q^2<(m_B-m_\pi)^2$, is indispensable for the determination of
the CKM parameter $|V_{ub}|$ from the measurements of $B\to \pi\ell\nu_\ell$
decay \cite{delAmoSanchez:2010af,delAmoSanchez:2010zd,Ha:2010rf,Lees:2012vv,Sibidanov:2013rkk}. The
tension between this determination and the one
from inclusive $B\to X_u\ell\nu_\ell$ decays \cite{Beringer:1900zz} remains unresolved.
On the theory side,
the lattice QCD predictions for $B\to \pi$ form factors
are available at large momentum transfer \cite{Dalgic:2006dt,Bailey:2008wp,AlHaydari:2009zr}.
The method of QCD light-cone sum rules (LCSR)
provides these form factors at
low and intermediate momentum transfer, $q^2\leq 12-15$ GeV$^2$;
for the most recent results see
\cite{Khodjamirian:2011ub, Bharucha:2012wy}.
Hence, it is very important to have reliable analytical tools
for extrapolating the form factors from the LCSR to the lattice QCD
region of $q^2$
and vice versa. These extrapolations and a reliable assessment of their accuracy
are important  for both the lattice QCD and the LCSR methods for several reasons:
to go beyond the region of their respective applicability; to check their mutual consistence;
and to confront the future more accurate results on the shape of the form factors.

For momentum transfer extrapolations, one usually considers a form factor
as an analytical (meromorphic) function of the variable $q^2$. A conformal mapping from the variable $q^2$
to the new variable $z$ is applied,
so that the complex $q^2$ plane is mapped onto the unit disc $|z| \leq 1$ in the complex $z$ plane. For $f^+_{B\pi}(q^2)$,
the semileptonic region $0<q^2<(m_B-m_{\pi})^2$
is then transformed to the interval of real
and small $z$, whereas the region of timelike momentum transfer
$(m_B+m_{\pi})^2<q^2<\infty $
is projected onto the unit circle.
Owing to  the smallness of $z$ in the semileptonic region, one  uses a Taylor expansion
for the form factors near $z=0$, truncating it to a certain maximal order.
The most convenient and reliable ansatz commonly used
is the BCL version \cite{Bourrely:2008za} of the $z$ parametrization.

One purpose of this paper is to assess
the overall accuracy of this particular parametrization of the form factors.
In what follows we will consider only
the vector form factor $f^+_{B\pi}(q^2)$ of the $B\to \pi$
transition for which the available
QCD calculations are most accurate to date. A similar analysis of other form factors
is deferred to a future work.
We will extensively use the LCSR calculations of $B\to \pi$
form factors carried out in \cite{Duplancic:2008ix,Khodjamirian:2011ub}
in the \MSbar{} scheme of the $b$-quark mass. The earlier results presented in \cite{Ball:2004ye}
as well as the most recent calculation
\cite{Bharucha:2012wy} use the pole-mass scheme and lead to numerically close results.

An important aspect, which so far has been missing in the previous extrapolations
of LCSR results from small to large $q^2$,
is the correlation between the uncertainties of normalization
and shape of  the form factor. These correlations arise from the simultaneous variations
of common input parameters.
In earlier analyses uncorrelated
uncertainties have been provided that underestimate the accuracy of the LCSR results.
To avoid the computation of uncertainty correlations, in \cite{Khodjamirian:2011ub}
the form factor squared was integrated over the LCSR region
and then used in the determination of $|V_{ub}|$.
However, this procedure leaves aside important aspects of the theory predictions,
such as the shape of the form factor that can be compared with anticipated
more accurate measurements of the semileptonic differential width. These measurements will provide
additional tests for the theoretical calculations.
In this paper, we carry out a comprehensive statistical analysis of the form factor uncertainties.
From variations of the input parameters in the LCSR
we infer theoretical uncertainties
of the fitted normalization and shape parameters of the $z$ parametrization,
including their correlations.

The second goal of this paper is to control the accuracy of the
extrapolation to larger momentum transfer. The widely used $z$ parametrization includes only the
first few terms of a Taylor expansion and is therefore model dependent.
We suggest to confront the extrapolation with
rigorous bounds that follow from the unitarity of the correlation function
of heavy-light vector currents,  and from the analyticity
of the form factors. We update the unitarity bounds for $B\to \pi$
form factors. Moreover, the bounds obtained using the LCSR inputs at a sufficiently large value of $q^2$
form a rather narrow band up to $q^2 \simeq 20\,\GeV^2$.
These limits become challenging to the extrapolation of the LCSR results and
to lattice QCD points.

The plan of the paper is as follows. In \refsec{calc} we
revisit the $B\to \pi$ vector form factor as calculated from
LCSR, and carry out the statistical analysis of its uncertainties. In \refsec{extrap}
we fit the BCL form of the $z$ parametrization to our LCSR results, and thus provide
an extrapolation to large momentum transfers.
Based on this extrapolation, we estimate the strong $B^*B\pi$ coupling in \refsec{bstarbpi}.
In \refsec{bounds} we obtain the unitarity bounds from the correlation
function, where our LCSR results for the form factor normalization, as well as its 
first and second derivative with respect to $q^2$ are used as inputs. The resulting bounds
are compared with the extrapolated BCL parametrization.
We further use our results in \refsec{vub} to determine $|V_{ub}|$ from various experimental results on the
exclusive decay $\bar{B}^0\to \pi^+\ell \bar\nu_\ell$.
Section \ref{sec:concl} contains the concluding discussion. For convenience, we
collect necessary elements of the unitarity bounds in the appendix.

\section{Calculation of the form factor from LCSR and uncertainty analysis \label{sec:calc}}

We use the standard definition of the $B\to \pi$ form factors
\begin{multline}
    \langle \pi^+(p)|\bar{u}
    \gamma_\mu   b |\bar{B}(p+q)\rangle\\
    = f^+_{B\pi}(q^2)\Big [2p_\mu +
    \big (1-\frac{m_B^2-m_\pi^2}{q^2}\big) q_\mu\Big]
    + f^0_{B\pi}(q^2)\frac{m_B^2-m_\pi^2}{q^2}q_\mu\,,
    \label{eq:Bpihme}
\end{multline}
where, for definiteness, we only consider the transition $\bar{B}^0\to \pi^+$.
It is our intent to set up the following statistical analysis of the vector form factor
$f_{B\pi}^+(q^2)$. The scalar form factor $f_{B\pi}^0(q^2)$, which is of secondary interest, will
not be analysed here.

Before turning to the numerical computation,
let us shortly outline the derivation
of the LCSR. The method was introduced in \cite{Balitsky:1986st,Braun:1988qv,Chernyak:1990ag} and
the first applications to $B\to \pi$ transitions go back to \cite{Belyaev:1993wp,Belyaev:1994zk,Khodjamirian:1997ub,Bagan:1997bp}.
For a detailed description of the LCSR for the form factor
$f_{B\pi}^+(q^2)$  we refer to \cite{Duplancic:2008ix} where all definitions and
resulting analytic expressions relevant to our numerical analysis are presented.

One starts from
the vacuum-to-pion correlation function of two $b$-flavoured quark currents:
\begin{multline} \label{eq:corr}
    i\!\! \int \!d^4xe^{iqx}\langle \pi^+(p)\!\mid T\{\bar{u}\gamma_\mu b(x),
    m_b\bar{b}i\gamma_5 d(0)\}\mid\! 0\rangle \\
    \equiv F((p+q)^2,q^2)p_\mu+ \widetilde{F}((p+q)^2,q^2)q_\mu\,,
\end{multline}
where the invariant amplitude $F((p+q)^2,q^2)$, considered as an analytical
function of the variable $(p+q)^2$,
is used to access the vector form factor $f^+_{B\pi}(q^2)$ at fixed $q^2$.
The second amplitude $\widetilde{F}$ is only needed for the calculation of
the scalar form factor $f_{B\pi}^0(q^2)$.

At $q^2\ll m_b^2$ and $(p+q)^2\ll m_b^2$ the intermediate $b$ quark
that propagates between the points $x$ and $0$
is highly virtual. The $T$-product of quark currents can therefore be expanded near the light-cone
$x^2\sim 0$. In the resulting operator-product expansion (OPE) the
$b$-quark fields are contracted to
a propagator and form perturbatively calculable
coefficient functions, whereas the light-quark fields are included in
universal  vacuum-pion
matrix elements of the type
$\langle \pi(p)|\bar{u}_\alpha(x) d_\beta(0)|0\rangle$
and $\langle \pi(p)|\bar{u}_\alpha(x)G_{\mu\nu}(vx)d_\beta(0)|0\rangle$.
They absorb nonperturbative effects and are expressed in terms of the pion distribution amplitudes (DAs)
with increasing twist. The light-cone OPE yields for the invariant amplitude
a generic decomposition:
\begin{equation}
    \label{eq:ope1}
    F^{(OPE)}((p+q)^2,q^2)=\sum\limits_{t=2,3,4,...}{\cal F}^{(t)}((p+q)^2,q^2) \,,
\end{equation}
where each separate twist component
\begin{multline}
    \label{eq:opet}
    {\cal F }^{(t)}((p+q)^2,q^2)\\
    = f_\pi\int \rmDx{u}
    \!\!\! \sum\limits_{k=0,1,...}
    \!\!\! \left(\frac{\alpha_s(\mu)}{\pi}\right)^k
    \! T^{(t)}_k((p+q)^2,q^2, \lbrace u_i\rbrace, \mu,m_b)
    \, \varphi^{(t)}_\pi(\lbrace u_i\rbrace,\mu; \vec\theta_{\text{DA}}^{\,(t)})\,,
\end{multline}
factorizes into a perturbatively expandable coefficient function $T_k^{(t)}$ and the pion
DAs $\varphi^{(t)}_\pi$ of twist $t$. The DAs depend on the specific set of input
parameters $\vec\theta_{\text{DA}}^{\,(t)}$.
In the above, $\mu$ is the factorization scale and we use the same
scale for renormalization of the quark-gluon coupling, $b$-quark mass
and all other running parameters in the adopted \MSbar{}
scheme. The integration variables $\lbrace u_i\rbrace = \lbrace u_1,u_2,...\rbrace$ correspond
to the fractions of the pion momentum carried by the quark and antiquark in the two-particle Fock state, and the
quark, antiquark and gluon in the three-particle Fock state:
$\rmDx{u} \equiv \delta\left(1 - \sum_i u_i\right) \prod_i \rmdx{u_i}$.

The terms in \refeq{ope1} that correspond to higher-twist
pion DAs are suppressed by inverse powers of the $b$-quark
virtuality $ ((p+q)^2-m_b^2)\sim \bar{\Lambda}m_b $,
where $\bar{\Lambda}\gg \Lambda_{QCD}$ does not scale with $m_b$.
This allows for truncation of the OPE.
We use the same approximation for the correlation function
as in \cite{Duplancic:2008ix}, which includes: all LO
contributions of the twist 2,3,4 quark-antiquark and
quark-antiquark-gluon DA's of the pion; and the
NLO, $O(\alpha_s)$ corrections to the twist-2 and twist-3
two-particle coefficient functions. We do not include
the recently calculated $\beta_0$ estimation for the twist-2 $O(\alpha_s^2)$ contributions \cite{Bharucha:2012wy},
since the resulting effect is very small and
does not yet represent a complete NNLO calculation.
Note that the $u,d$ quark masses and the pion mass are neglected with one exception, the ``chirally enhanced''
parameter $\mu_\pi \equiv m_\pi^2/(m_u+m_d)$, which appears in the normalization of twist-3 DAs.

The input parameters of the pion DAs, normalized to the
pion decay constant $f_\pi$, are specified at a fixed normalization scale,
typically at $\mu_0 = 1\,\GeV$, with a logarithmic renormalization group evolution to the scale $\mu$.
For the twist-2 pion DA these are the coefficients $a_n^\pi$ of the expansion in terms of Gegenbauer
polynomial. We adopt a usual ansatz with two nonvanishing coefficients $a^\pi_2$ and $a^\pi_4$, so that $a^\pi_{>4}=0$.
For the higher twist DAs the conformal expansion  is truncated at the first nonleading order (see the analysis in \cite{Ball:2006wn}).
The pion DAs of nonleading twist are slightly updated with respect to \cite{Duplancic:2008ix}; the definitions we use here can be found in \cite{Ball:2006wn}
(see also \cite{Khodjamirian:2009ys}).
Beside $\mu_\pi$, the twist-3 components of the OPE include two further parameters:
the normalization of the three-particle DA $f_{3\pi}$ and the coefficient
of the nonasymptotic part $\omega_{3\pi}$.
Finally we use two additional parameters, $\delta_\pi^2$ and $\omega_{4\pi}$,  for the normalization and nonasymptotic parts of the twist-4 DAs,
respectively.
To summarize, the set of pion DA parameters is:
\begin{equation}
\begin{aligned}
    \vec\theta_{\text{DA}}^{\,(2)} & = (a_2^\pi, a_4^\pi),\nonumber \\
    \vec\theta_{DA}^{\,(3)} & = (\mu_\pi(2\,\GeV), f_{3\pi}(1\,\GeV), \omega_{3\pi}(1\,\GeV)),\nonumber\\
    \vec\theta_{DA}^{\,(4)} & = (\delta^2_\pi(1\,\GeV), \omega_{4\pi}(1\,\GeV)).
\end{aligned}
\label{eq:thetaDA}
\end{equation}

In order to link the correlation function to the $B \to \pi$ form factor,
we use the  hadronic dispersion relation for the invariant amplitude in the variable $(p+q)^2$,
but at a fixed value of $q^2$,
\begin{equation}
    F((p+q)^2,q^2)
    =\frac{2m_B^2f_B f^+_{B\pi}(q^2)}{m_B^2-(p+q)^2}
    + ...\,.
\label{eq:disp-F}
\end{equation}
The form factor enters
the ground-state $B$-meson contribution only as a product with the $B$-meson decay constant
$f_B$. The remaining sum over excited and continuum hadronic states
with quantum numbers of the $B$-meson, indicated by ellipses in \refeq{disp-F},
is approximated using (semi-local) quark-hadron duality: the sum
in the hadronic dispersion relation is replaced by the quark-gluon spectral density, obtained by
calculating $\Im F^{(OPE)}(s,q^2)$ from the OPE \refeq{ope1} at $(p+q)^2\equiv s>m_b^2$.
The duality approximation introduces the effective threshold parameter $s_0^B$.
After the Borel transformation $(p+q)^2 \to M^2\sim \bar{\Lambda}m_b$,
the resulting LCSR for the $B\to \pi$ form factor
has the following form
\begin{equation}
    f_{B\pi}^+(q^2; \vec\theta)
    = \left(\frac{e^{m_B^2/M^2}}{2m_B^2 [f_B]_\text{2ptSR}}\right)
        \int\limits_{m_b^2}^{s_0^B}\rmdx{s}\,
        \sum\limits_{t=2,3,4}\frac{1}{\pi}\Im {\cal F}^{(t)}(s,q^2)\,e^{-s/M^2}\,.
\label{eq:lcsr}
\end{equation}
Anticipating the following numerical and statistical analysis,
we introduce $\vec\theta$, the set of all input parameters on which
the r.h.s.\ of the LCSR depends. The components of $\vec\theta$ shall later be varied
within certain intervals. Furthermore, the label ``2ptSR'' at the decay constant $f_B$
in \refeq{lcsr} indicates that we
calculate it from the QCD sum rule for the two-point correlation function of $m_b\bar{b}i\gamma_5 d$ currents.
Without going into details, which can be found e.g.\ in the recent update
of this sum rule in \cite{Gelhausen:2013wia}, we write the result of this calculation schematically as
\begin{equation}
    \left[f_B^2\right]_{\text{2ptSR}}
    = \left(\frac{e^{m_B^2/\overline{M}^2}}{m_B^4}\right)\overline{\cal F}(\overline{M}^2, \overline{s}^B_0, \alpha_s, \mu, m_b, \vec\theta_\text{cond})\,.
\label{eq:fB}
\end{equation}
Here, the quantity $\overline{\cal F}$ represents the result of the OPE
for the two-point correlation function. It contains perturbative
and nonperturbative contributions; the latter involve vacuum condensate densities of growing mass dimension.
In this sum rule, owing to the power suppression of the terms
with higher-dimensional condensates, only terms up to the mass dimension six are taken into account in the OPE.
The corresponding set of input parameters for the vacuum condensates
in the adopted approximation,
\begin{equation}
    \vec\theta_\text{cond} = \left(\langle\bar{q}q\rangle(2\,\GeV), \langle\frac{\alpha_s}{\pi} G^2\rangle, m_0^2, r_\text{vac}\right)\,,
\end{equation}
includes the quark-condensate density at the reference scale,
the (scale independent) gluon-condensate density,
the ratio of quark-gluon to quark-condensate density
$m_0^2 \equiv \langle \bar{q}G\bar{q}\rangle / \langle \bar{q}q\rangle$,
and the coefficient $r_\text{vac}$ that parametrizes the factorization
in the four-quark condensate density. We neglect a weak scale dependence of the
higher-dimensional condensate densities. Note that we use the two-point sum rule
to NLO accuracy, to stay consistent with the overall accuracy of the LCSR.

Summarizing the input parameters in \refeq{lcsr}, where
eqs.~(\ref{eq:fB}) and (\ref{eq:opet}) are substituted, we have
for the set of variable inputs:
\begin{equation}
    \vec\theta 
    \equiv \left(\alpha_s(M_Z)\,,\overline{m}_b(m_b)\,,\vec\theta_{DA}^{\,(2,3,4)}\,,\vec\theta_\text{cond}\,,M^2,\,s_0^B\,,\overline{M}^2\,,\overline{s}_0^B\right) \,.
\label{eq:theta}
\end{equation}
We exclude from this set the meson masses
$m_B$, $m_\pi$ and the pion decay constant $f_\pi$ that are all measured with a negligibly small
errors. We also exclude the combined renormalization and factorization scale $\mu$.
This follows after explicitly checking that variation of $\mu$
in the preferred interval $[2.5\,\GeV,\,4\,\GeV]$, chosen according to \cite{Khodjamirian:2011ub},
yields shifts to the mean values which are small
compared to the remaining parametric uncertainties.
Within the perturbative expansion of the invariant amplitudes $F^\text{(OPE)}$,
\begin{equation}
    F^\text{(OPE)} = \sum_{k=0,1,2,\dots} \left(\frac{\alpha_s C_F}{4 \pi}\right)^k F^{(k)}\,,
\end{equation}
we also investigate the effects of hypothetical NNLO terms $F^{(2)}$. We estimate the latter as
\begin{equation}
    |F^{(2)}| \leq \left|\frac{[F^{(1)}]^2}{F^{(0)}}\right|\,,
\end{equation}
and find the associated uncertainty to be well below $0.05\%$ for $0 \leq q^2 \leq 12\,\GeV^2$.

In addition, the derivatives of the LCSR and of the two-point sum rule
with respect to $-1/M^2$ and $-1/\overline{M}^2$, respectively, should reproduce
the $B$-meson mass squared:
\begin{equation}
\begin{aligned}%
    \,[m_B^2]_{LCSR} & = \frac{\int\limits_{m_b^2}^{s_0^B} ds s
    \mbox{Im}F^{(OPE)}(s,q^2)e^{-s/M^2}}{\int\limits_{m_b^2}^{s_0^B} ds
    \mbox{Im}F^{(OPE)}(s,q^2)e^{-s/M^2}}\,,\\
    [m_B^2]_\text{2ptSR} &= \frac{ d/d(-1/\overline{M}^2)
    \overline{\cal F}(\overline{M}^2, \overline{s}_0^B, \dots)}{
    \overline{\cal F}(\overline{M}^2, \overline{s}_0^B, \dots)}\,.
\end{aligned}
\label{eq:mB}
\end{equation}
We will use these relations later on in our statistical analysis, where we constrain both
$[m_B]_{\text{LCSR}}$ and $[m_B]_{\text{2ptSR}}$ to lie within a given small interval around
the measured $B$-meson mass. This is a more general constraint than the one used in the previous
analyses of these sum rules, such as in \cite{Duplancic:2008ix,Khodjamirian:2011ub}.
There, the threshold parameters were adjusted to the $B$-meson mass by variation of only the
Borel parameter.

\begin{table}[t]
\centering
\renewcommand{\arraystretch}{1.2}
\resizebox{\textwidth}{!}{
\begin{tabular}{|c| c| c| c| c|}
    \hline
    Parameter                                   & value/interval                 & unit     & prior               & source/comments\\
    \hline
    \multicolumn{5}{|c|}{quark-gluon coupling and quark masses}\\
    \hline
    $\alpha_s(m_Z)$                             &  0.1184  $\pm$ 0.0007          & ---      & gaussian $@$ $68\%$ &  \cite{Beringer:1900zz}                \\
    $\mbMSbar$                                  &    4.18  $\pm$ 0.03            & \GeV     & gaussian $@$ $68\%$ &  \cite{Beringer:1900zz} \\
    $m_s$                                       &  95 $\pm$ 10                   & \MeV     & ---                 & \cite{Beringer:1900zz} (error doubled)\\
    $R\equiv 2m_s/(m_u+m_d)$                    &  24.4 $\pm$ 1.5                &---       & ---                 &   ChPT, \cite{Leutwyler:1996qg}\\
    $m_u+m_d$                                   & 7.8 $\pm$ 0.9                  & \MeV     & gaussian $@$ $68\%$ & $2m_s/R$ \\
    \hline
    \multicolumn{5}{|c|}{hadron masses}\\
    \hline
    $m_B$                                       &  5279.58                       & \MeV     & ---                 &  \cite{Beringer:1900zz}                \\
    $m_\pi$                                     & 139.57                         & \MeV     & ---                 &  \cite{Beringer:1900zz}                \\
    \hline
    \multicolumn{5}{|c|}{vacuum condensate densities}\\
    \hline
    $\langle\bar{q}q (2\GeV)\rangle$            & $-(277^{+12}_{-10})^3$         & $\MeV^3$ & ---                 &  $m_\pi^2f_\pi^2/2(m_u+m_d)$\\
    $\langle\frac{\alpha_s}{\pi} G^2\rangle$    & $[0.000, 0.018]$               & $\GeV^4$ & uniform $@$ $100\%$ &  \cite{Ioffe:2002ee}              \\
    $m_0^2$                                     & $[0.6, 1.0]$                   & $\GeV^2$ & uniform $@$ $100\%$ &   \cite{Ioffe:2002ee}                         \\
    $r_{vac}$                                   & $[0.1, 1.0]$                   & ---      & uniform $@$ $100\%$ &  \cite{Ioffe:2002ee}                   \\
    \hline
    \multicolumn{5}{|c|}{parameters of the pion DAs}\\
    \hline
    $f_\pi$                                     & $130.4$                        & \MeV     & ---                 &  \cite{Beringer:1900zz}                \\
    $a_{2\pi}(1 \GeV)$                          & $[0.09, 0.25]$                 & ---      & uniform $@$ $100\%$ &  \cite{Khodjamirian:2011ub}               \\
    $a_{4\pi}(1 \GeV)$                          & $[-0.04, 0.16]$                & ---      & uniform $@$ $100\%$ &   \cite{Khodjamirian:2011ub}          \\
    $\mu_\pi(2 \GeV)$                           & 2.5 $\pm$ 0.3                  & \GeV     & ---                 &  $m_\pi^2/(m_u+m_d)$ \\
    $f_{3\pi}(1 \GeV)$                          & $[0.003, 0.006]$               & $\GeV^2$ & uniform $@$ $100\%$ & \cite{Ball:2006wn}                     \\
    $\omega_{3\pi}(1 \GeV)$                     & $[-2.2, -0.8]$                 & ---      & uniform $@$ $100\%$ &  \cite{Ball:2006wn}                    \\
    $\delta_\pi^2(1 \GeV)$                      & $[0.12, 0.24]$                 & $\GeV^2$ & uniform $@$ $100\%$ & \cite{Ball:2006wn}                     \\
    $\omega_{4\pi}(1 \GeV)$                     & $[0.1, 0.3]$                   & ---      & uniform $@$ $100\%$ & \cite{Ball:2006wn}                        \\
    \hline
    \multicolumn{5}{|c|}{sum rule parameters and scales}\\
\hline
    $\mu$                                       & $3.0$                          & $\GeV$   & ---                 &  \cite{Khodjamirian:2011ub,Gelhausen:2013wia}                   \\
    $M^2$                                       & $16.0 \pm 4.0$                 & $\GeV^2$ & gaussian $@$ $68\%$ & \cite{Khodjamirian:2011ub}                \\
    $s_0^B$                                     & $[30.0, 45.0]$                 & $\GeV^2$ & uniform $@$ $100\%$ &                 \\
    $\overline{M}^2$                            &  $5.5 \pm 1.0$                 & $\GeV^2$ & gaussian $@$ $68\%$ &   \cite{Gelhausen:2013wia}                    \\
    $\overline{s}_0^B$                          & $[29.0, 44.0]$                 & $\GeV^2$ & uniform $@$ $100\%$ &                 \\
    \hline
\end{tabular}
}
\caption{Input parameters used in the numerical analysis.
The prior distribution $P_0(\vec\theta)$ is a product of individual priors, either
uniform or gaussian. The uniform priors cover the stated intervals with 100\% probability.
The gaussian priors cover the stated intervals with 68\% probability, and the central value
corresponds to the mode of the prior. For practical purposes, variates from the gaussian
priors are only drawn from their respective 99\% intervals.
}
\label{tab:inputs}
\end{table}

In \reftab{inputs} we collect all input parameters and their
adopted variation ranges
used in our numerical calculation, indicating also their sources.
A few comments are in order.
Note that we use the strange quark mass determination
\footnote{We double the error to be consistent with the typical
accuracy of non-lattice determinations of $m_s$.}
from \cite{Beringer:1900zz} in a combination with the very
accurate ChPT relation \cite{Leutwyler:1996qg}
to determine $m_u+m_d$ that in turn yields the quark-condensate
density and the parameter $\mu_\pi$.
The intervals for $a^\pi_2,a^\pi_4$ were estimated in \cite{Khodjamirian:2011ub}
by fitting the LCSR for the pion
electromagnetic form factor to the experimental data.
Our choice of the pion twist-2 DA is consistent with the
two-point sum rule estimates of $a_2^\pi$. Moreover, as shown in \cite{Khodjamirian:2011ub}
the pion DA with four nonvanishing Gegenbauer moments,
based on the updated analysis of the LCSR
for the photon-pion transition form factor \cite{Agaev:2010aq}
(see also \cite{Agaev:2012tm}), produces very similar results for the
$B\to \pi$ form factor. Furthermore, we specify the
uniform renormalization and factorization scale $\mu$
as well as the Borel-parameter intervals for both LCSR and 2-point sum rule
according to the choice in \cite{Khodjamirian:2011ub} and \cite{Gelhausen:2013wia}.
The broad ranges for both threshold parameters
will be substantially constrained by the $B$-meson mass relations.
In our numerical analysis, we use four-loop running of $\alpha_s$
and quark masses, whereas for the nonperturbative scale-dependent
parameters of the pion DA's one-loop (LL) renormalization suffices.

Important is that there are in fact more correlations
between many of the input parameters, apart from the one
between the quark-condensate density, $\mu_\pi$ and light-quark masses
that is taken into account.
For example, the normalization
and nonasymptotic coefficients of higher twist pion DAs
are themselves obtained from two-point sum rules; i.e., they depend
on the condensate parameters. However, the task of including all
these ``hidden'' correlations remains outside the scope of our present work.
It demands a global simultaneous numerical analysis
of all sum rules involved in the determination of the input parameters.
We make here the simplifying assumption that all parameters entering
$\vec\theta$ are independent and their individual uncertainties are therefore not correlated.
Due to this conservative assumption, we expect that the uncertainties
estimated in this paper are in fact somewhat larger than the true ones.

We now turn to the details of the statistical analysis.
Throughout this work we use a Bayesian approach (see e.g. \cite{BayesianDataAnalysis} for a review) to determine the
mean values and theoretical uncertainties for the form factor $f^+_{B\pi}(q^2)$.
To this end, we have implemented the two-point sum rule for $f_B$, as well
as the LCSR for $f_{B\pi}^+$ within EOS \cite{EOS}, a HEP program for the computation
of flavour observables. Throughout this work, all numerical results are obtained using EOS.
We start from the region of momentum transfer where the LCSR is applicable.
The upper limit of this region is chosen to be $q^2=12\,\GeV^2$ as in \cite{Khodjamirian:2011ub},
to guarantee a good convergence of the light-cone OPE.
This is reflected by the very small size of the twist-4 contribution to the LCSR.
We express our prior knowledge of the input parameters $\vec\theta$ through the
\emph{prior distribution} $P_0(\vec\theta)$, for which we use a product of uncorrelated
distributions for each element of $\vec\theta$.
The individual factors are either uniform or
gaussian distributions. The uniform distributions cover the intervals listed
in \reftab{inputs} at $100\%$ probability. The gaussian ones use the listed
intervals so that the central value is the mode, and the intervals
contain $68\%$ of accumulated probability. However, the allowed interval for such parameters
is restricted to their respective $99\%$ probability intervals for practical purposes.

We now construct a \emph{likelihood} $P(m_B|\vec\theta)$ that incorporates \emph{purely theoretical}
constraints on our parameter space. Specifically, we demand that
theory determinations of the $B$-meson mass $[m_B]_{\text{LCSR}}(q^2)$ and $[m_B]_{\text{2ptSR}}$
from \refeq{mB} agree with the experimentally measured value within $1\%$
at $68\%$ probability. It should be emphasized that the magnitude of $\sigma_B$ stems purely from
our estimates.
We thus use a gaussian distribution for the likelihood with the standard
deviation parameter $\sigma_B = m_B \cdot 1\% \simeq 0.053\,\GeV$.
The complete likelihood reads
\begin{equation}
    P(m_B|\vec\theta)
    = \mathcal{N}(m_B, \sigma_B; [m_B]_\text{2ptSR}(\vec\theta))
    \times \mathcal{N}(m_B, \sigma_B; [m_B]_\text{LCSR}(\vec\theta;q^2 = 0))\,,
\end{equation}
where $\mathcal{N}(\mu, \sigma; x)$ denotes the probability density function
for the gaussian distribution of $x$ around the mean $\mu$ with standard deviation $\sigma$.
From our prior and likelihood follows the \emph{posterior distribution} according to Bayes' theorem,
\begin{equation}
    P(\vec\theta|m_B) = \frac{P(m_B|\vec\theta) P_0(\vec\theta)}{\int \mathrm{d}\vec\theta\,P(m_B|\vec\theta) P_0(\vec\theta)}\,.
\end{equation}
We observe that the likelihood has significant impact on the posterior distribution of the threshold
parameters $s_0^B$ and $\overline{s}_0^B$. Their marginalized one-dimensional posteriors resemble a gaussian
distribution, with approximate $68\%$ intervals of $(41\pm 4)\GeV^2$
and $(35\pm 2)\GeV^2$, respectively. The remainder of the input parameters
exhibit virtually no difference between their respective priors and marginalized one-dimensional posteriors.
While \refeq{mB} at $q^2 = 0$ perfectly reproduces the experimental $B$-meson mass, we observe
that the mode of the distribution of $[m_B]_\text{LCSR}(q^2 = 10\,\GeV^2)$ is
shifted with respect to $m_B$ by $\sim 2.5\%$.

In order to obtain the values and uncertainties for the form factor results from our
input parameters, we carry out an uncertainty propagation.
This is achieved by computing $5\cdot 10^4$ samples of the joint \emph{posterior predictive distribution} $P(\vec{F}|m_B)$,
\begin{equation}
    P(\vec{F}|m_B) = \int \mathrm{d}\vec\theta \delta(\vec{F} - \vec{F}(\vec\theta)) P(\vec\theta|m_B)\,,
    \label{eq:post-pred-dist}
\end{equation}
of our quantities of interest $\vec{F}$. For the latter we chose the normalization of the form factor
as well as its first and second derivative with respect to $q^2$. We evaluate these quantities
at two points $q^2=0$ and $q^2=10 \,\GeV^2$ located at the opposite ends of the LCSR region:
\begin{equation}
    \vec{F} \equiv \left(f_{B\pi}^+(0),\,f_{B\pi}^{+\prime}(0),\, f_{B\pi}^{+\prime\prime}(0),\,
f_{B\pi}^+(10\,\GeV^2),\, f_{B\pi}^{+\prime}(10\, \GeV^2),\, f_{B\pi}^{+\prime\prime}(10\,\GeV^2)\right)\,.
\end{equation}
In the above, we denote the first and second derivative of $f_{B\pi}^+(q^2)$ with respect to $q^2$ as
$f_{B\pi}^{+\prime}(q^2)$ and $f_{B\pi}^{+\prime\prime}(q^2)$, respectively.

We find that the one-dimensional marginal distributions for each element of $\vec F$
resemble a gaussian distributions to good accuracy. The one-dimensional marginalised
posterior of $f_{B\pi}^{+\prime\prime}$ is slightly leptokurtic, with $\operatorname{Kurt} \sim 4$.
The remainder of the one-dimensional posteriors are approximately mesokurtic, with $|\operatorname{Kurt}| < 0.25$.
We feel therefore confident to approximate the true distribution $P(\vec{F}|m_B)$ through a six-dimensional multivariate
gaussian distributions, $P(\vec{F}|m_B) \simeq \mathcal{N}_6(\vec\mu^F,\Sigma^F;\vec{F})$. We obtain the
mean vector $\vec{\mu}^F$, the vector of the standard deviations $\vec\sigma^F$, and the correlation
matrix $\rho^F$ as
\begin{gather}
    \label{eq:LCSRmean}
    \vec\mu^F
    = \left(0.310,\,1.55\cdot10^{-2},\,1.24\cdot10^{-3},\,0.562,\,4.03\cdot10^{-2},\,4.71\cdot10^{-3}\right),\\
    \label{eq:LCSRstddev}
    \vec\sigma^F
    = \left(0.020,\,0.10\cdot10^{-2},\,0.10\cdot10^{-3},\,0.032,\,0.24\cdot10^{-2},\,0.37\cdot10^{-3}\right),\\
    \label{eq:LCSRcorr}
    \rho^F
    =
    \left(
    \renewcommand{\arraystretch}{1.33}
    \begin{array}{cccccc}
        1.000 & 0.735 & 0.374 & 0.925 & 0.564 & 0.313 \\
        0.735 & 1.000 & 0.867 & 0.927 & 0.863 & 0.246 \\
        0.374 & 0.867 & 1.000 & 0.682 & 0.853 & 0.221 \\
        0,925 & 0.927 & 0.682 & 1.000 & 0.814 & 0.389 \\
        0.564 & 0.863 & 0.853 & 0.814 & 1.000 & 0.647 \\
        0.313 & 0.246 & 0.221 & 0.389 & 0.647 & 1.000 \\
    \end{array}
    \renewcommand{\arraystretch}{1.0}
    \right)\,.
\end{gather}
The covariance matrix is then $\Sigma^F_{ij} \equiv \sigma^F_i \sigma^F_j \rho^F_{ij}$.

Note that we obtain an uncertainty on $f_{B\pi}^+(0)$ that is about $20\%$ smaller than the uncertainty
given in \cite{Khodjamirian:2011ub}, where similar ranges for the numerical inputs have been used. We
expect that further improvement on the precision of $f_{B\pi}^+(0)$ can be achieved if the input parameters
can be further constrained. Inclusion of two-point sum rules for the determination of $a_\pi^2$ and $a_\pi^4$,
and of experimental measurements of the pion electromagnetic form factors as part of the likelihood are
good candidates for such an improvement. Moreover, fits to our LCSR
results for the $B\to \pi$ vector form factor will benefit from the correlation
matrix $\rho^F$, which is computed for the first time.

Our result $f^+_{B\pi}(0) = 0.31 \pm 0.02$ is somewhat larger
than the determination in \cite{Khodjamirian:2011ub}, with the value
$f^+_{B\pi}(0)=0.281$ obtained for the central input.
Roughly half of the change is due to a slight update of the
input parameters, such as the $b$-quark mass in the \MSbar{} scheme.
We have explicitly checked that our normalization with nominal input
parameters as in \cite{Khodjamirian:2011ub} reproduces their results.
The remainder of the change is due to the different statistical treatment,
since we quote the mode of the posterior predictive distribution.
On the other hand, our result and the one in \cite{Khodjamirian:2011ub} exhibit
a tension with the calculation \cite{Ball:2004ye} (and the partial NNLO update
\cite{Bharucha:2012wy}).
The latter results, obtained in the pole scheme for the $b$-quark mass, exhibit
a smaller central value of $f^+_{B\pi}(0)$. For a detailed discussion
on the difference between the respective calculational procedures we refer
to \cite{Duplancic:2008ix}.

For completeness we also calculate the integral $\Delta\zeta(0,12\GeV^2)$,
\begin{equation}
    \Delta\zeta\,(0, 12 \GeV^2)
    \equiv \frac{G_F^2}{24\pi^3}\int\limits_0^{12\,\text{GeV}^2}dq^2p_\pi^3
    |f_{B\pi}^+(q^2)|^2 = (5.25^{+0.68}_{-0.54})\,\text{ps}^{-1}\,,
\label{eq:zeta}
\end{equation}
where $p_\pi=\sqrt{(m_B^2+m_\pi^2-q^2)^2/4m_B^2-m_\pi^2}$ is the pion's
spatial momentum in the $B$-meson rest frame. The integration range of $\Delta\zeta$
covers the adopted domain of validity for the LCSR.
Our $\Delta\zeta$ is compatible with the results of \cite{Khodjamirian:2011ub}, with a relative
increase of $14\%$. This larger central value is consistent with the increase
by $8\%$ in the normalization of the central value of the form factor.

\section{Extrapolation of the form factor toward large momentum transfer \label{sec:extrap}}

Having calculated the form factor in the LCSR region, we now turn
to an extrapolation toward large momentum transfer $q^2$. To this end, we
employ a $z$-series parametrization where we transform the $q^2$-variable
in the standard way:
\begin{equation}
z(q^2, t_0) =
\frac{\sqrt{t_+ - q^2}-\sqrt{t_+-t_0}}{\sqrt{t_+-q^2}+\sqrt{t_+-t_0}}\,.
\label{eq:zparam}
\end{equation}
Here and throughout we use $t_{\pm} \equiv (m_B\pm m_\pi)^2$ and
adopt $t_0$ following \cite{Bourrely:2008za},
\begin{equation}
    t_{0,\text{opt}} = (m_B + m_\pi) \cdot (\sqrt{m_B} - \sqrt{m_\pi})^2 \simeq 20\,\GeV^2\,,
\label{eq:t0opt}
\end{equation}
so that $|z|$ is sufficiently small in the semileptonic region $0<q^2<t_{-}$,
\begin{equation}
    |z(q^2, t_{0,\text{opt}})| < 0.280\,,
\end{equation}
and $z(q^2, t_0)$ is positive in the domain of validity of the LCSR.

In what follows, we use the $K=3$ form of the z-series expansion \cite{Bourrely:2008za},
modified to use the form factor at $q^2=0$ as one of the parameters. Is it
given by the three-parameter expression:
\begin{eqnarray}
    f^+_{B\pi}(q^2) = \frac{f^+_{B\pi}(0)}{1 - q^2/m_{B^*}^2}
\Bigg\{1 + b_1^+ \Bigg[z(q^2,t_0) - z(0,t_0)
- \frac{1}{3} \Big( z(q^2,t_0)^3 - z(0,t_0)^3\Big)\Bigg]
\nonumber \\
    + b_2^+ \Bigg[z(q^2,t_0)^2 - z(0,t_0)^2 + \frac{2}{3}\Big(z(q^2,t_0)^3
- z(0,t_0)^3\Big)\Bigg]\Bigg\}\,.
\label{eq:BCL}
\end{eqnarray}
The form factor is parametrized in terms of $f_{B\pi}^+(0)$,
as well as the two shape parameters $b_1^+$ and $b_2^+$.
An analogous but somewhat simpler expression with only one shape parameter
was used in previous works \cite{Khodjamirian:2009ys,Khodjamirian:2011ub}.

\begin{figure}[t]
    \centering
    \begin{tabular}{cc}
        \includegraphics[width=.4\textwidth]{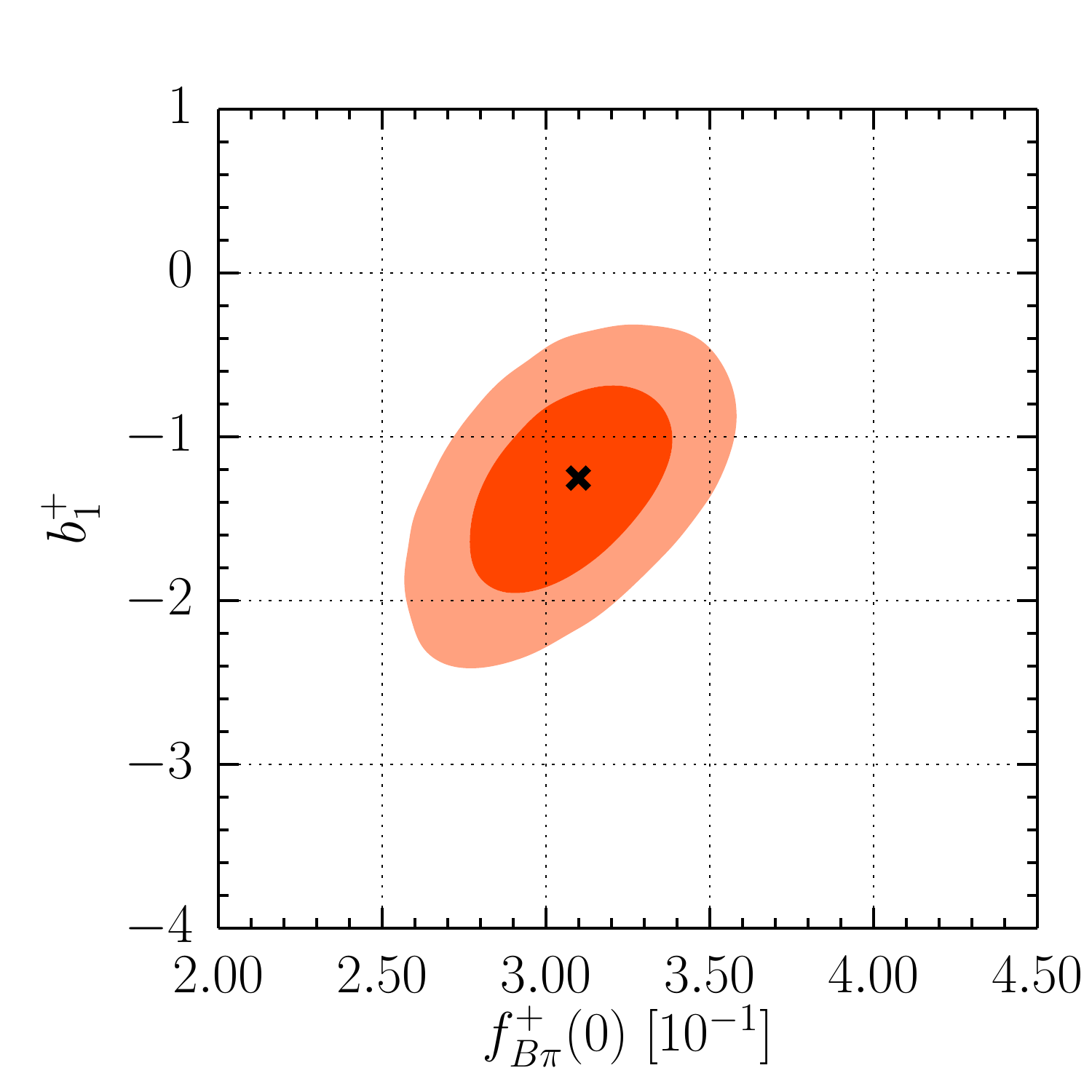} &
        \includegraphics[width=.4\textwidth]{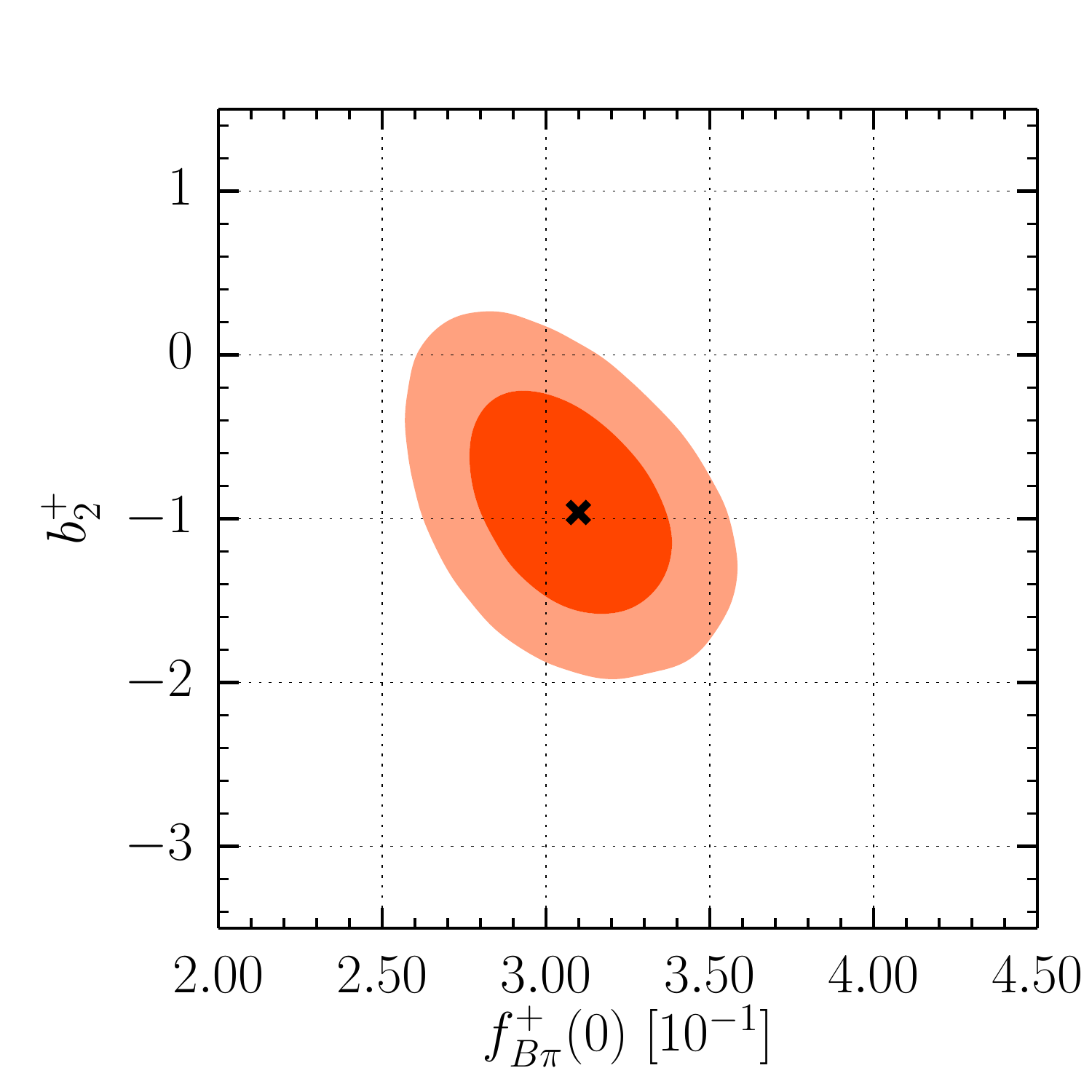} \\
        \includegraphics[width=.4\textwidth]{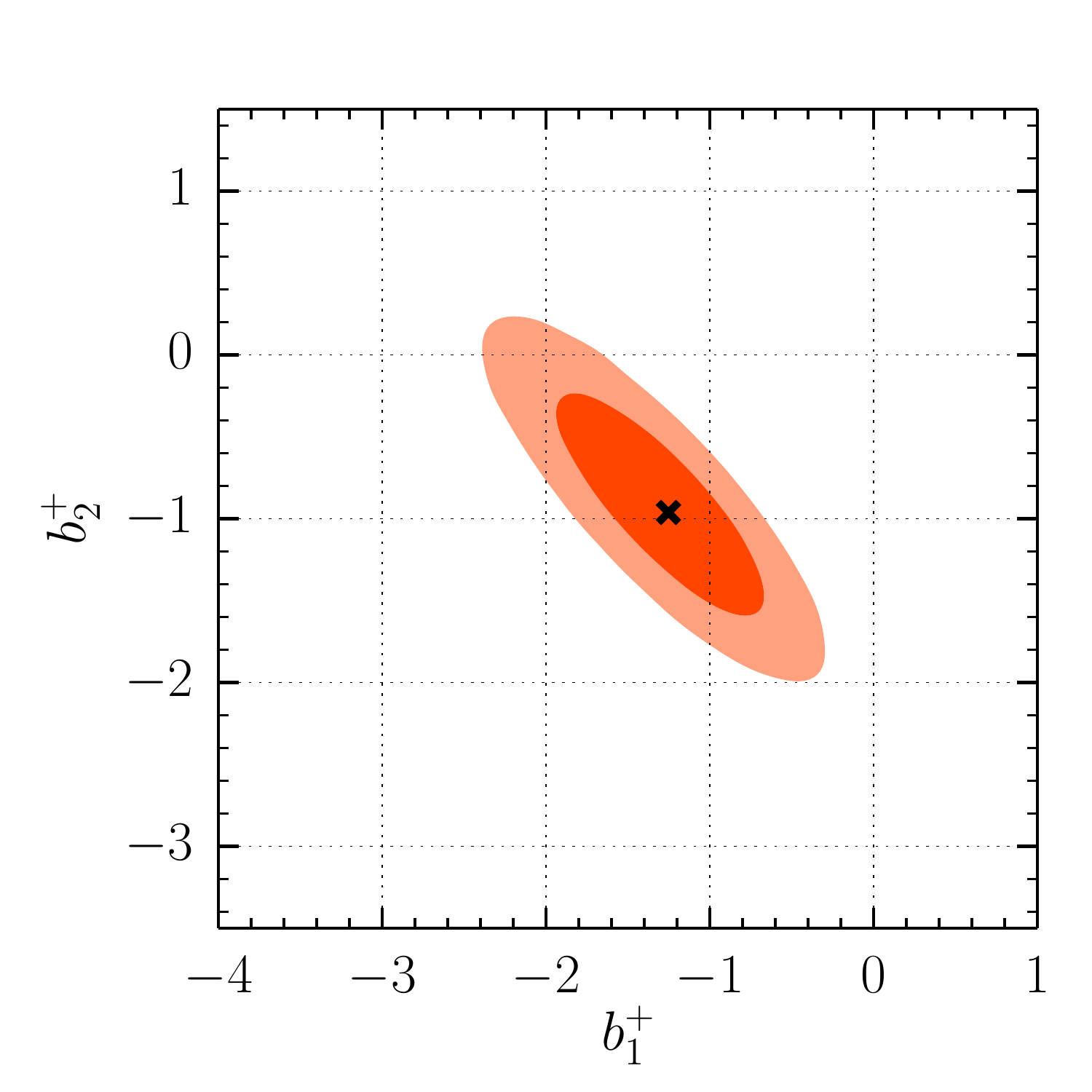} &
        \begin{minipage}[b][.4\textwidth][t]{.4\textwidth}
        \caption{The regions with $68\%$ probability (red) and $95\%$ probability (orange)
        for all two-dimensional marginalisations of the posterior $P(\vec\lambda|\text{LCSR})$.
        The cross marks the best-fit point.
        \label{fig:BCL}
        }
        \end{minipage}
    \end{tabular}
\end{figure}

We fit the parametrization \refeq{BCL} to our LCSR results, which are incorporated in the likelihood
\begin{equation}
    \label{eq:LHLCSR}
    P(\text{LCSR}|\vec\lambda) \equiv \mathcal{N}_6(\vec\mu^F,\Sigma^F; \vec{F}(\vec\lambda))\,,
\end{equation}
with $\vec\mu^F$ and $\Sigma^F$ as determined in \refsec{calc}; see eqs. (\ref{eq:LCSRmean})--(\ref{eq:LCSRcorr}). We choose the prior $P_0(\vec\lambda)$ as uniform for
all three BCL parameters $\vec\lambda = (f_{B\pi}^+(0), b_1^+, b_2^+)$,
with the respective support intervals
\begin{equation}
    \begin{aligned}
        0 & \leq f_{B\pi}^+(0) \leq 1\,,  &    -10 & \leq b_1^+ \leq +10\,, &    -10 & \leq b_2^+ \leq +10\,.
    \end{aligned}
\end{equation}
\begin{figure*}[t]
    \centering
    \includegraphics[width=0.75\textwidth]{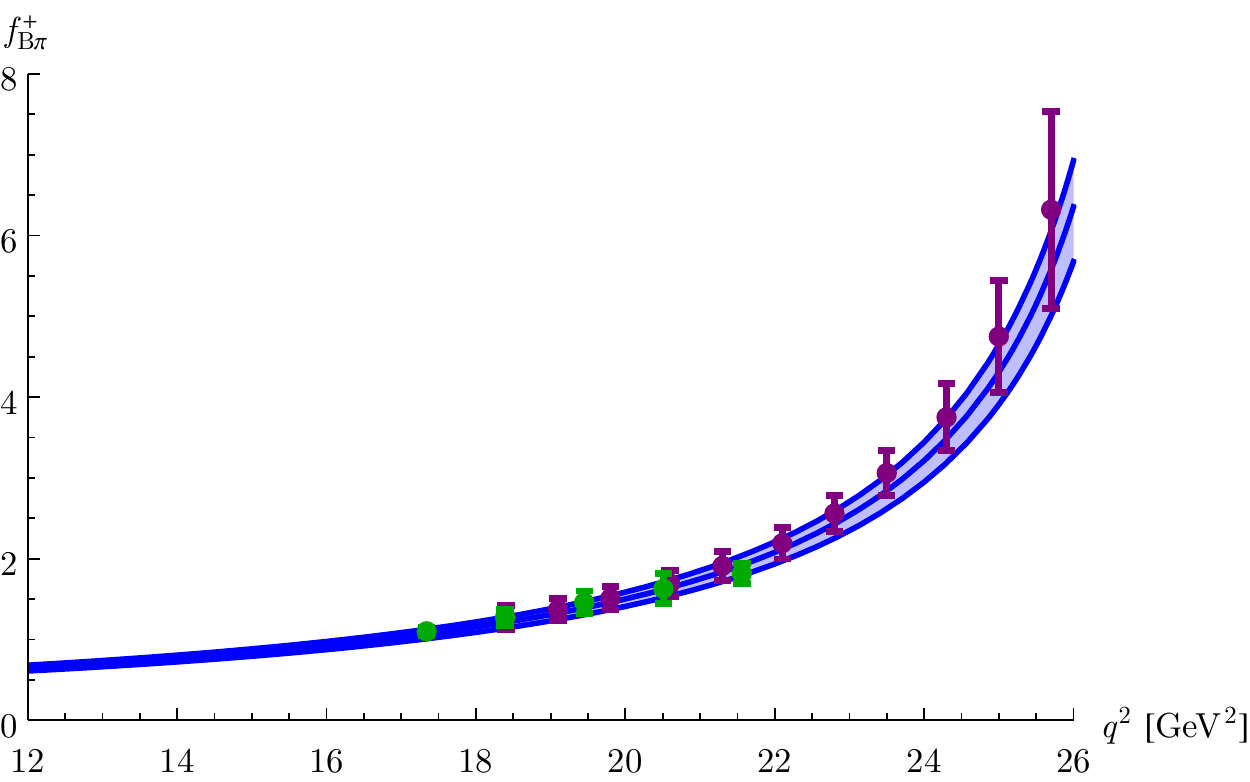}
    \caption{Form factor $f^+_{B\pi}(q^2)$ obtained
    at $q^2<12\GeV^2$ from the statistical analysis of LCSR, fitted to z-series representation and extrapolated
    to large $q^2$. The solid lines correspond to the $68\%$ probability envelope and the best fit curve.
    The green (magenta) points are HPQCD \cite{Dalgic:2006dt} (Fermilab-MILC \cite{Bailey:2008wp}) lattice
    QCD results.
    }
    \label{fig:BCLvslatt}
\end{figure*}
From this follows the posterior distribution $P(\vec\lambda | \text{LCSR})$,
which turns out to be gaussian to very good approximation.
This is evidenced by very small skewness $|\operatorname{Skew}| < 0.2$, and
kurtosis $|\operatorname{Kurt}| < 0.2$. We therefore choose to approximate
the posterior as
\begin{equation}
    P(\vec\lambda |\text{LCSR}) \simeq \mathcal{N}_3(\mu^\text{BCL}, \Sigma^\text{BCL}; \vec\lambda)\,,
\end{equation}
where
\begin{gather}
    \vec\mu^\text{BCL}    = (0.307, -1.31, -0.904)\,,\\
    \vec\sigma^\text{BCL} = (0.020,  0.42,  0.444)\,,\\
    \rho^\text{BCL}       =
    \left(
    \begin{array}{ccc}
      1.000 &  0.503 & -0.391 \\
      0.503 &  1.000 & -0.824 \\
     -0.391 & -0.824 &  1.000 \\
    \end{array}
    \right)\,,
\end{gather}
and with $\Sigma^\text{BCL} \equiv \sigma^\text{BCL}_i \sigma^\text{BCL}_j \rho^\text{BCL}_{ij}$.
\footnote{%
    In addition, a data file containing $10^5$ weighted variates of the
    posterior can be obtained from the authors upon request.
}
For the best-fit point we obtain
\begin{equation}
    \label{eq:BCLbfp}
    \vec\lambda^*_\text{LCSR} \equiv \argmax P(\vec\lambda|\text{LCSR}) = (0.310, -1.25, -0.962)\,.
\end{equation}
The goodness of fit is excellent, with $\chi^2 = 1\cdot 10^{-7}$. For this fit
we count our predictions as six observations. Our three fit parameters thus reduce
the number of degrees of freedom down to $N_\text{d.o.f.} = 3$.
Using the corresponding $\chi^2$ distribution, we estimate a p value of $>0.99$.
For completeness, we show the $68\%$ and $95\%$ contours for all two-dimensional
marginalisations of the posterior in \reffig{BCL}.

We compare the extrapolation of our LCSR results of $f_{B\pi}^+$ to large $q^2$ with
lattice data points from the HPQCD \cite{Dalgic:2006dt} and the Fermilab-MILC \cite{Bailey:2008wp}
collaborations in \reffig{BCLvslatt}. There, the filled region corresponds to
possible values of the form factor at $68\%$ probability.
We also compute the pull values of the best-fit extrapolation with the respective data points in
\reftab{lattice-pulls},
\begin{equation}
    \operatorname{pull} \equiv \frac{[f^+_{B\pi}(q^2)]_\text{lattice} - [f^+_{B\pi}(q^2)]_\text{extrap.}}{\sigma_\text{lattice}}\,,
\end{equation}
where $\sigma_\text{lattice}$ denotes the uncertainty for the respective lattice point.
Without information on the correlation among the lattice results,
we cannot compute their goodness of fit with respect to our best-fit-point. However, we do compute
the pull values on a point-by-point basis, and find no pull in excess of $0.83\sigma$.
We consider this a good agreement between our extrapolation
and the lattice results.
Furthermore, we observe that, except for one HPQCD lattice point, all lattice points listed in \reftab{lattice-pulls} exhibit positive pull values, which indicates that the
lattice results are systematically larger than favoured by our extrapolation.
This is of special interest in light
of recent preliminary lattice QCD results \cite{Bouchard:CKM2014}, which are smaller than the published
results listed in \reftab{lattice-pulls}.
The observed trend toward positive pull values could, for instance, be explained by different shapes of the form factors
as computed using LCSR and lattice QCD. However, in absence of correlation information for the lattice QCD results,
a more detailed comparison of the shapes is not yet possible.

\begin{table}[t]
\begin{center}
    \renewcommand{\arraystretch}{1.2}
    \begin{tabular}{|c|c|c|c|c|c|}
        \hline
        $q^2$     & $f_{B\pi}^+(q^2)$ & Pull     & $q^2$     & $f_{B\pi}^+(q^2)$ & Pull      \\
        \hline
        \multicolumn{3}{|c|}{Fermilab-MILC}      & \multicolumn{3}{|c|}{Fermilab-MILC}       \\
        \hline
        $18.4$    & $1.27$            & $+0.39$  & $25.0$    & $4.75$            & $+0.64$   \\
        $19.1$    & $1.37$            & $+0.32$  & $25.7$    & $6.32$            & $+0.61$   \\
        $19.8$    & $1.51$            & $+0.35$  & $26.5$    & $9.04$            & $+0.36$   \\
                                                 \cline{4-6}
        $20.6$    & $1.69$            & $+0.31$  & \multicolumn{3}{|c|}{HPQCD}               \\
                                                 \cline{4-6}
        $21.3$    & $1.91$            & $+0.41$  & $17.34$   & $1.101$           & $+0.58$   \\
        $22.1$    & $2.19$            & $+0.38$  & $18.39$   & $1.273$           & $+0.62$   \\
        $22.8$    & $2.56$            & $+0.58$  & $19.45$   & $1.458$           & $+0.48$   \\
        $23.5$    & $3.06$            & $+0.79$  & $20.51$   & $1.627$           & $+0.04$   \\
        $24.3$    & $3.75$            & $+0.64$  & $21.56$   & $1.816$           & $-0.83$   \\
        \hline
    \end{tabular}
    \renewcommand{\arraystretch}{1.0}
\end{center}
\caption{The pull values of lattice data points from Fermilab-MILC \cite{Bailey:2008wp} and HPQCD \cite{Dalgic:2006dt}
with respect to our $z$-series extrapolation of the LCSR results. The extrapolation
is obtained from the best-fit point $\lambda^*$, see text. The sum of the $\chi^2$ values are $1.72$
and $3.67$ for HPQCD and Fermilab-MILC, respectively.}
\label{tab:lattice-pulls}
\end{table}

\section{Estimation of the strong $B^*B\pi$ coupling \label{sec:bstarbpi}}

It is well known that the $B\to \pi$ form factor obeys a rigorous hadronic dispersion
relation without subtractions, which reads
\begin{equation}
f^+_{B\pi}(q^2)= \frac{g_{B^*B\pi}f_{B^*}}{2m_{B^*}(1-q^2/m_{B^*}^2)}+
\frac1\pi\int\limits_{t_+}^\infty dt\frac{\mbox{Im} f_{B\pi}^+(t)}{t-q^2}
\label{eq:disp}\,.
\end{equation}
Here, the residue of the  $B^*$-pole located below threshold $t_+$
is proportional to the strong $B^*B\pi$ coupling, and to the $B^*$ decay constant that
is defined as $\langle 0|\bar{u} \gamma^\mu b|B^*(p, \epsilon_{B^*})\rangle = f_{B^*}m_{B^*}\epsilon_{B^*}^\mu$.
For the decay constant we use
\begin{equation}
    f_{B^*}=210 \pm 11 \, \MeV\,,
\label{eq:fBst}
\end{equation}
which was recently obtained from a 2-point QCD sum rule with NNLO accuracy \cite{Gelhausen:2013wia}.
It is instructive to compare our above extrapolation
to large $q^2$ with this dispersion relation. First of all,
owing to the fact that the $B^*$ pole is embedded in the BCL-ansatz,
we can directly estimate the strong coupling. It can be obtained as the residue of the $B^*$-pole
from \refeq{BCL}:
\begin{equation}
    g_{B^*B\pi} =
    \frac{2 m_{B^*}}{f_{B^*}} \lim_{q^2 \to m_{B^*}^2} \Big[(1- q^2/m_{B^*}^2) f_{B\pi}^+(q^2)\Big]\,.
\end{equation}
We find
\begin{equation}
    g_{B^*B\pi} = 30 \pm 5\,,
    \label{eq:BstBpi}
\end{equation}
at $68\%$ probability.
Note that since this estimate is obtained through an extrapolation
slightly beyond the semileptonic region, an additional
"systematic" error related to the truncation of $z$ series
is expected, according to \cite{Bourrely:2008za}.
The interval \refeq{BstBpi} is in the same ballpark as the first LCSR determination
of this hadronic matrix element \cite{Belyaev:1994zk},
see also \cite{Khodjamirian:1999hb, Ball:2004ye}.
An update of the LCSR for $g_{B^*B\pi}$ with a better precision than
our extrapolation-based result would therefore be useful to anchor the BCL parametrization at
large $q^2$-values.

Lattice QCD results on the $B^*B\pi$ coupling
are usually presented
in a form of the effective coupling $g_b$ in the heavy-meson chiral
perturbation theory; at leading order the relation is
\begin{equation}
    \frac{2m_B}{f_\pi} g_b = g_{B^*B\pi}\,,
\end{equation}
so that from (\ref{eq:BstBpi}) we obtain  $g_b=0.35\pm 0.06$.
This value is compatible with, but smaller than, most lattice QCD results,
e.g. $g_b=0.449\pm 0.047 \pm 0.019$ \cite{Detmold:2012ge}.
Since our extrapolation (\ref{eq:BstBpi}) estimates
the strong $B^*B\pi$ coupling in full QCD at finite $b$ quark mass,
the issue of inverse heavy-mass corrections is important for this
comparison, but remains outside the scope of our present work.

Returning to the dispersion relation (\ref{eq:disp})
and substituting the residue of the $B^*$ pole,
we are now in a position to estimate the integral on r.h.s.\ at various $q^2$, assessing the
cumulative contribution to $f_{B\pi}^+(q^2)$ of
radially excited and continuum states
with $B^*$ quantum numbers.
It is easy to notice that at $q^2=0$ this contribution
is comparable with the one from $B^*$-pole but has an opposite sign.

\section{Unitarity bounds with inputs from LCSR\label{sec:bounds}}

The $z$-parametrization of the form factor $f^+_{B\pi}(q^2)$
obtained from LCSR results has a truncated form
depending on the maximal power chosen in the Taylor series at $z=0$.
It is important to test the extrapolations beyond the LCSR region
that are based on this parametrization.
To this end, we suggest to use the upper and lower bounds
on the form factor $f^+_{B\pi}(q^2)$ that are obtained from the unitarity
of the correlation function of the $b$-flavoured vector currents,
combined with the input from the LCSR calculation.
Since the unitarity bounds are valid in the whole semileptonic region,
they are independent of any parametrization of the form factor.

For $B\to \pi$ form factors the unitarity bounds have been derived in
\cite{Lellouch:1995yv}, following an earlier work \cite{Bourrely:1980gp}.
One starts from the two-point correlation function:
\begin{eqnarray}
\label{eq:ope}
\Pi_{\mu\nu}(q) = i\int\! d^4x\, e^{iq\cdot x} \langle 0 | T\left\{
\bar{u}\gamma_\mu b(x), \bar{b}\gamma_\nu u (0) \right\} | 0 \rangle
\nonumber \\
= ( - g_{\mu\nu}q^2+q_\mu q_\nu) \widetilde{\Pi}_T(q^2) + q_\mu q_\nu \Pi_L(q^2)\,.
\end{eqnarray}
In what follows, we only
need the transverse invariant amplitude $\widetilde{\Pi}_T(q^2)$,
multiplying it by $q^2$ to avoid kinematical singularities.
It is calculated at $q^2\ll m_b^2$ using an OPE in terms of the same
condensate densities as used in the two-point sum rule \refeq{fB}.
We update the calculation of \refeq{ope} with respect to the previous
analyses of the bounds, adding the NNLO, $O(\alpha_s^2)$ correction
to the perturbative part and the NLO, $O(\alpha_s)$ correction to the quark-condensate part. To this end, we use  the results of \cite{Gelhausen:2013wia} where the same correlation function was employed to obtain the QCD sum rule for the decay constant of the $B^*$ meson.
The input parameters
needed to calculate (\ref{eq:ope}) are the same as for the correlation function of the pseudoscalar $b$-flavoured currents used for the $f_B$ calculation and are already
specified in table \ref{tab:inputs}.

As a further step we match the OPE result to the hadronic dispersion relation
and subsequently differentiate both sides of this relation $n+1$ times with respect to $q^2$ at $q^2=\tilde{q}^2
\ll m_b^2$:
\ba
\chi_T^{\text{(OPE)}}(\tilde{q}^2,n) \equiv
\frac{1}{(n+1)!}\left(\frac{\partial}{\partial q^2}\right)^{n+1}
\big[q^2\widetilde{\Pi}^\text{(OPE)}_T(q^2)\big]\Bigg|_{q^2=\tilde{q}^2}
= \int\limits_{m_{B^*}^2}^\infty dt\,\frac{ \rho_T(t) }{(t-\tilde{q}^2)^{n+2}}\,,
\label{eq:chin}
\ea
where $\rho_T(t)\equiv \frac{1}{\pi}{\rm Im} [t \widetilde{\Pi}_T(t)] $
is the hadronic spectral density.
Note that the dispersion integral converges
for $n\geq 1$ as follows from QCD asymptotics of the perturbative part of the correlation function.
At $n=1$, \refeq{chin} coincides with the standard double-subtracted dispersion relation used in \cite{Lellouch:1995yv}. According to the unitarity condition,
$\rho_T(s)$ contains positive contributions of all possible hadronic states with the $B^*$ quantum numbers. 
The ground $B^*$-state  contribution to the r.h.s. of differentiated dispersion
relation has a form:
\begin{equation}
\chi_T^{(B^*)}(\tilde{q}^2,n) = \frac{f^2_{B^*} m_{B^*} ^2}{(m_{B^*}^2-\tilde{q}^2)^{n+2}}\,,
\label{eqn:chiBstar}
\end{equation}
where for the decay constant $f_{B^*}$ we use the interval \refeq{fBst}.
The hadronic continuum state $|B\pi\rangle $ in $P$-wave, with
the threshold $t_+$ located slightly above $m_{B^*}^2$,
interests us because its contribution to \refeq{chin}
contains the integral over the $B\to \pi$ vector form factor squared:
\be
\chi_T^{(B\pi)}(\tilde{q}^2,n)= \int_{t_+}^\infty dt\, k_T(t,\tilde{q}^2,n)|f^+_{B\pi}(t)|^2,
\label{eq:fBpicontrib}
\ee
where the function
\begin{eqnarray}
k_T(s,\tilde{q}^2,n) &=& \frac{1}{32\pi^2}\frac{\left[(t-t_+)(t-t_-)\right]^{3/2}}{t^2(t-\tilde{q}^2)^{n+2}}\,,
\label{eq:kT}
\end{eqnarray}
contains all kinematical factors.
In the above, we also take into account the isospin weights of  the
two (equal in the isospin limit)  contributions $\bar{B}^0\pi^-$ and $B^-\pi^0$
with $b\bar{u} $ quantum numbers.
A straightforward upper limit for the integrated form factor squared follows then 
from \refeq{chin}:
\begin{equation}
    \chi_T^{(B\pi)}(\tilde{q}^2,n) \leq  \chi^{\text{(OPE)}}_T(\tilde{q}^2,n)- \chi^{(B^*)}_T(\tilde{q}^2,n)\,.
\label{eq:boundn}
\end{equation}
In \cite{ Bourrely:1980gp,Lellouch:1995yv} and in later analyses
$n=1$, $\tilde{q}^2=0$ was chosen in the above.
We have carried out a detailed investigation, in which we vary $n$ and take also negative
values of $\tilde{q}^2$. The resulting bounds do not improve upon such a variation. Hence, in
what follows we only consider
the simplest case $n=1$, $\tilde{q}^2=0$
and abbreviate:
\begin{equation}
    \chi^+\equiv \chi^{\text{(OPE)}}_T(0,1)- \chi^{(B^*)}_T(0,1)\,,
\label{eq:chi}
\end{equation}
so that
\be
\chi_T^{(B\pi)}(0,1) \leq \chi^+\,.
\label{eq:corrbound}
\ee
To proceed to a more elaborated bound
that also includes the above upper limit, we
map the momentum transfer variable $q^2 \to z(q^2,t_0)$
according to \refeq{zparam}. As a consequence,
the region of integration in \refeq{fBpicontrib}
transforms into the unit circle in the $z$ plane.
Note that we are free to choose an arbitrary, convenient 
value of the parameter $t_0$ in this transformation; i.e., our choice
of $t_0$ need not be the same as in \refsec{extrap}, \refeq{t0opt}.
In terms of the  new variable,  eq.(\ref{eq:corrbound}) takes the form
\begin{eqnarray}
\label{eq:correlbound}
 \int\limits_{|z|=1} \frac{dz}{2\pi i z} \left|\phi(z,0,1)f_{B\pi}^+(t(z,t_0))\right|^2
\leq \chi^+\,,
\end{eqnarray}
where $t(z)$ is the inverse of the transformation \refeq{zparam}.
Hereafter we suppress  the fixed parameter $t_0$  in
the argument of $z$-variable  for brevity. In the above, $\phi(z,\tilde{q}^2,n)$,
given in Appendix,
is the outer function, which is analytic and without zeros at
$|z|<1$, and whose modulus on the boundary $|z|=1$
is related to the kinematical weight function \refeq{kT} and the Jacobian of the
transformation \refeq{zparam}.

Furthermore,
one introduces a general definition of the inner product
of two functions, integrating the product of $f_1(z)$ and
the complex conjugate of $f_2(z)$  on the unit circle:
\begin{eqnarray}\label{eq:ip1}
\langle f_2|f_1 \rangle = \int_{|z|=1} \frac{dz}{2\pi i z} f_2^*(z)f_1(z)\,.
\end{eqnarray}
In terms of this definition \refeq{correlbound}
can be rewritten as
\begin{eqnarray}
\label{eq:ip2}
\langle \phi f_{B\pi}^+|\phi f_{B\pi}^+\rangle \equiv\langle h|h\rangle
\leq \chi^+\,,
\end{eqnarray}
where we denote:
\begin{equation}
h(z)\equiv\phi(z,0,1) f^+_{B\pi}(q^2(z))\,.
\label{eq:hz}
\end{equation}
Note that our choice of the outer function contains
    the Blaschke factor, which effectively removes the $B^*$ pole;
    see the appendix. As a consequence, the function $h(z)$ is analytic
    at $|z| < 1$, and real-valued for $z \in (-1, +1)$.
\begin{figure*}[t]
    \centering
    \subfigure[]{\includegraphics[width=0.75\textwidth]{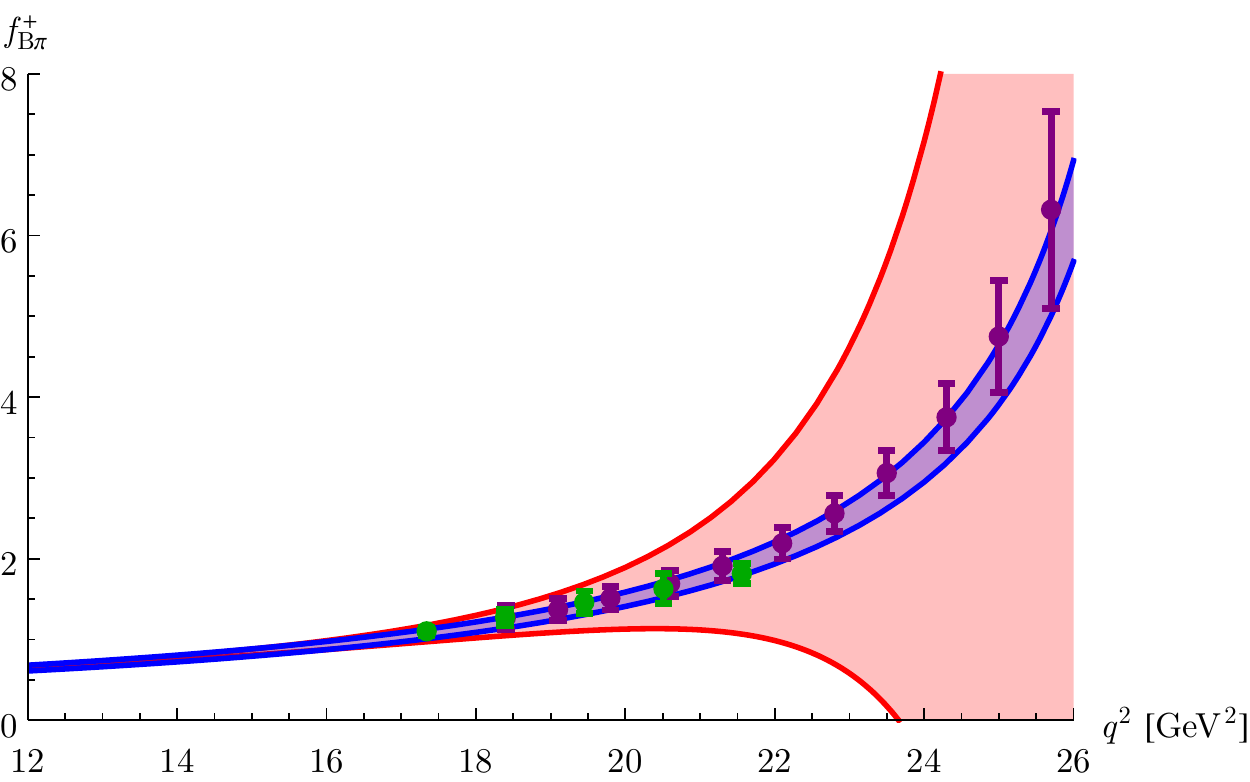}}\\
    \subfigure[\label{fig:bounds:zoom}]{\includegraphics[width=0.75\textwidth]{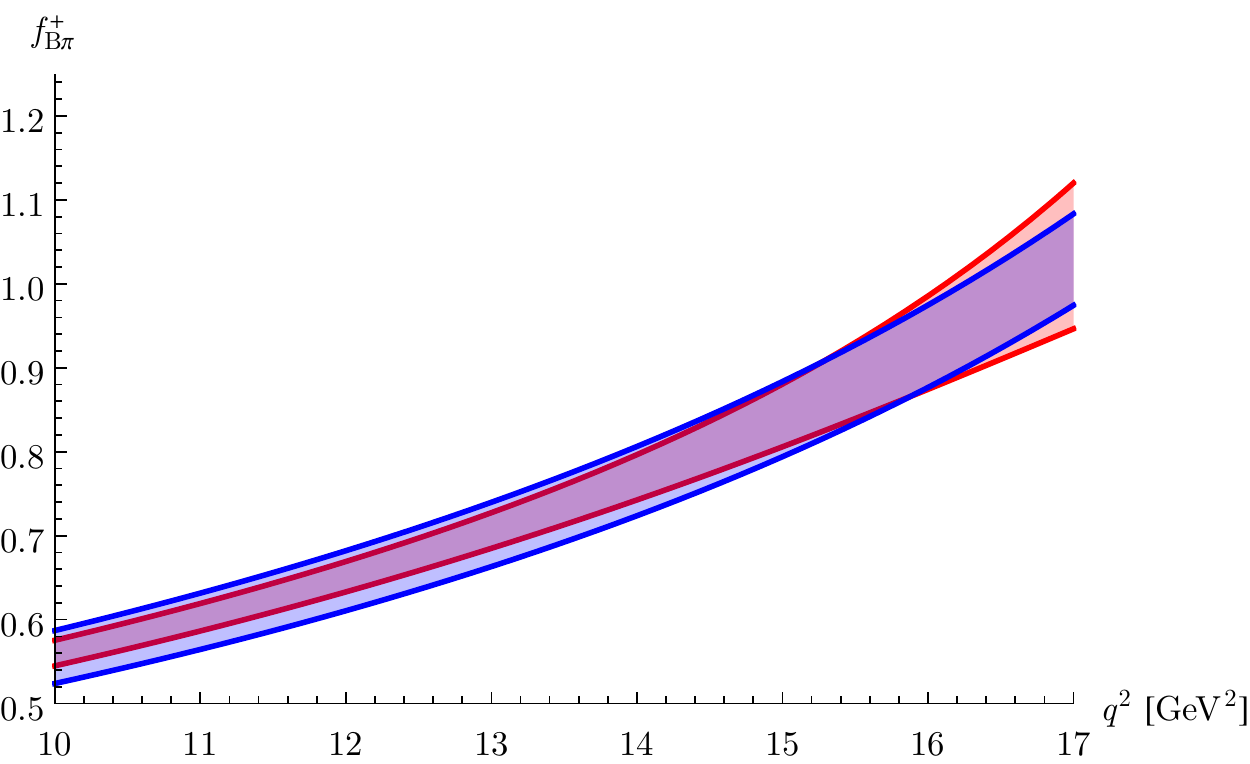}}
    \caption{Unitarity bounds (red shaded area) based on the LCSR input at $q^2=10\,\GeV^2$
        in comparison with the lattice QCD results (green points: HPQCD \cite{Dalgic:2006dt},
        magenta points: Fermilab-MILC \cite{Bailey:2008wp}) and our extrapolation based on the
        BCL parametrization (blue shaded area).}
    \label{fig:bounds}
\end{figure*}
Employing the Cauchy theorem with the circle $|z|=1$ taken as a contour
in the complex $z$-plane, one finds the value of this function
at any point $z(q^2)$ of the real $z$-axis  in terms of the inner product:
\begin{equation}
h(z(q^2))=\langle g| h \rangle\,, 
\label{eq:cauchy}
\end{equation}
where
\begin{equation}
\label{eq:ip3}
g(z(q^2),z(t)) = \frac{1}{1-z^*(q^2)z(t)}\,.
\end{equation}
In the above, we distinguish the $z$-values on the real axis
and on the unit circle by labeling their arguments as
$q^2$ (in the semileptonic region) and $t$ (at $t\geq t_+$), respectively.

In \cite{Lellouch:1995yv}, the lattice QCD values of the form factor were
employed as an additional input, leading to a substantial
improvement of the resulting bounds at large $q^2$.
Accordingly, we can adopt as an input the values of the form factor as
obtained from LCSR at an accessible small and intermediate values of
$q^2$. More specifically,
in  our analysis we follow \cite{Mannel:1998kp}, where it was
noticed that it is more effective to use as an input
the value, first and second derivative
of the form factor calculated at one given value $q^2_0$; i.e.,
$f_{B\pi}^+(q_0^2)$, $f^{+'}_{B\pi}(q_0^2)$ and
$f^{+''}_{B\pi}(q_0^2)$, respectively.\footnote{%
    A different approach to include the values of the form factors
    and their higher derivatives is based \cite{Abbas:2010jc} on
    Lagrange multipliers and analytic interpolation theory, and leads to similar results.
}
Hence, we will use the results of the LCSR obtained
at $q_0^2=10$ GeV$^2$ and presented in section 2.
Having at hand these three LCSR inputs  and the explicit
form of the outer function presented in the appendix, we calculate
at $z_0\equiv z(q_0^2)$ the function \refeq{hz}
and its first and second derivative. We abbreviate
\begin{equation}
h(z_0)\equiv h_0, ~~ \frac{dh(z)}{dz}\Big|_{z=z_0}\equiv h_1,~~  \frac12 \frac{d^2h(z)}{dz^2}\Big|_{z=z_0}\equiv h_2\,.
\label{eq:h012}
\end{equation} 
Simultaneously, we adopt $t_0=q_0^2$ so that $z_0=0$.

For the sake of generality we prefer to keep $z_0\neq 0$ and consider three points on the real $z$ axis: $z_0$, $z_0+\epsilon$
and  $z_0+\epsilon$
where $\epsilon$ is an arbitrarily small interval, so that
\be
h(z_0\pm \epsilon)=  h_0\pm h_1\epsilon +h_2 \epsilon^2+O(\epsilon^3)\,.
\label{eq:hexp}
\ee
The values  of $h(z_0)$ and  $h(z_0\pm \epsilon)$ (the latter
with $O(\epsilon^2)$ accuracy) can be transformed to the form of inner products similar to
\refeq{cauchy}:
\begin{equation}
h(z_0)=\langle g_0 | h\rangle\,,\quad
h(z_0 +\epsilon)=\langle g_{+} | h\rangle\,,\quad
h(z_0-\epsilon)=\langle g_{-} | h\rangle\,,
\label{eq:g1}
\end{equation}
where $g_0 \equiv g(z_0,z(t))$ and $g_\pm \equiv g(z_0 \pm \epsilon, z(t))$.

The next step is to form a $5\times 5$ matrix ${\cal M}$ that is
obtained by combining all diagonal and nondiagonal  
inner products of the above three functions,
together with the functions  $g(z(q^2),z(t))$ and $h(z(q^2))$.
The latter depends on the form factor $f_{B\pi}^+(q^2)$
at a certain value $q^2$ within the semileptonic region.

The explicit form of this matrix, given in the appendix, is obtained
after applying Cauchy's theorem for each inner product, except for the first
diagonal matrix element $\langle h |h\rangle$, and by employing
the approximation \refeq{hexp}.
We then use the positivity of all matrix elements of ${\cal M}$ 
from which the  condition  $\det{\cal M}\geq 0$ follows.\footnote{The simplest
mathematical construction of this type, without
any inputs, leads to a  $2\times 2$ matrix, see e.g., \cite{Lellouch:1995yv}.}
The latter can be rewritten in the form of quadratic inequality with
respect to $h(z)$:
\begin{equation}
\langle h |h\rangle\, d + a\,h(z)^2+b\,h(z) +c \geq 0\,.
\label{eq:finbounds}
\end{equation}
The coefficients $a,b,c,d$ are functions of $\epsilon$ and the input parameters $h_{0,1,2}$.
Inspection of these coefficients shows that all of them are proportional to $\epsilon ^6$ 
and the coefficient $d$ is positive. Hence, one can replace the diagonal inner product
$\langle h|h\rangle$ in the above by the upper limit \refeq{ip2} and simultaneously rescale the coefficients.
The resulting inequality is valid at all (sufficiently small) $\epsilon$, and 
a smooth limit $\epsilon \to 0$ is applicable. The two roots of the quadratic inequality
yield the upper and lower limits for the function 
$h(z)$.
Application of (\ref{eq:hz}) allows us to convert the latter into upper and lower bounds on
the form factor $f_{B\pi}^+$.
The result is especially simple at our choice $q_0^2=t_0$ and $z_0=0$:
\begin{equation}
\bigg[f_{B\pi}^+(q^2(z))\bigg]^{\rm up}_{\rm low}=\frac{ 
1}{\phi(z,0,1)}\Bigg(h_0+ h_1 z+ h_2 z^2
\pm \sqrt{\frac{z^6}{1-z^2}\big(\chi^+-h_0^2-h_1^2-h_2^2\big)}\Bigg)\,.
\label{eq:bounds}
\end{equation}

For a numerical computation of the bounds we require the form factor and its first and
second derivative at $q^2 = q_0^2 = 10\,\GeV^2$:
\begin{equation*}
    \vec F_{10} \equiv (f_{B\pi}^+(10\,\GeV^2), f_{B\pi}^{+\prime}(10\,\GeV^2), f_{B\pi}^{+\prime\prime}(10\,\GeV^2))\,.
\end{equation*}
Through marginalisation of $P(\vec F|m_B)$ (see \refeq{post-pred-dist}), we obtain the
required posterior predictive distribution $P(\vec F_{10}|m_B)$,
which we then use to compute the predictive distributions for the bounds as functions of $q^2$.
For simplicity, the value of $\chi^+$ is computed only for
the central input from \reftab{inputs}, after having checked that a variation
of the differentiated two-point correlation function at $q^2=0$ produces
a negligible impact on the bounds.
We obtain samples of the predictive distributions for both the upper and the lower bounds,
and consequently their respective cumulative distributions. We proceed to compute the
values $f^+_{B\pi,\text{up(low)}}(q^2)$ at $68\%$ ($1 - 68\%$) cumulative probability,
which are displayed as the red-shaded area in \reffig{bounds}, together with the lattice QCD results
and our extrapolations based on the BCL parametrization. We observe that the bounds are obeyed
by the central value of our extrapolation. For $q^2 < 16\,\GeV^2$, the bounds are somewhat more constraining
than the $68\%$-probability envelope, see \reffig{bounds:zoom}. This is not unexpected due to the fact that
the bounds and the extrapolation follow from different statistical analyses.
Interestingly enough, the bounds are constraining the lattice results quite critically at $q^2<20\,\GeV^2$.

\section{Determination of $|V_{ub}|$ from $\bar{B}^0\to \pi^+\ell^-\bar\nu$ decays \label{sec:vub}}

As the final step of this work we apply our results from \refsec{calc} to a determination of $|V_{ub}|$
from $B\to \pi \ell \nu$ decays. For this, we carry out a combined Bayesian fit of both $|V_{ub}|$
and the BCL parametrization \refeq{BCL} to the LCSR results and corresponding experimental results
by the BaBar and Belle collaborations \cite{Lees:2012vv,Sibidanov:2013rkk,delAmoSanchez:2010zd,Ha:2010rf}.
To this end, we extend EOS \cite{EOS} by implementing the relevant branching ratio for $\bar{B}^0\to \pi^+\ell \bar\nu_\ell$
decays, as well as the experimental likelihoods. All estimation of probability regions
follow from the algorithm as developed in \cite{Beaujean:2013}, which is implemented within EOS.

Let $\vec{x} = (|V_{ub}|, f_{B\pi}^+(0), b_1^+, b_2^+)$ denote the fit parameters. We construct
the prior $P_0(\vec x)$ using uniform distributions with the support
\begin{equation}
    \begin{aligned}
        1 & \leq 10^3 |V_{ub}| \leq 6\,,  &    -10 & \leq b_1^+ \leq +10\,,\\
        0 & \leq f_{B\pi}^+(0) \leq 1\,,  &    -10 & \leq b_2^+ \leq +10\,.
    \end{aligned}
\end{equation}

\begin{table}[t]
    \centering
    \renewcommand{\arraystretch}{1.2}
    \begin{tabular}{|c|c|c|c|c|c|c|}
        \hline
                        & \multicolumn{3}{c|}{pull [$\sigma$]}    &           &         & \\
        Data set        & LCSR         & BaBar        & Belle     & $\chi^2$  & p value & $\log(Z)$\\
        \hline
        $D_\text{2010}$ & $0.71$       & $1.78$       & $1.38$    & $ 5.58$   & $0.98$  & $166.02$\\
        \hline
        $D_\text{2013}$ & $0.28$       & $3.07$       & $2.20$    & $14.34$   & $0.42$  & $155.48$\\
        \hline
    \end{tabular}
    \renewcommand{\arraystretch}{1.0}
    \caption{Goodness-of-fit quantities for both the ``2010'' and the ``2013'' data sets, assuming $N_\text{d.o.f} = 14$
    degrees of freedom. All pulls follow from a (multivariate) gaussian distribution. Here, $Z \equiv \int \rmdx{\vec{x}} P(\vec x|D) P_0(\vec{x})$ denotes the evidence.}
    \label{tab:gof}
\end{table}

We perform two individual fits to the data sets ``2010'' and ``2013'',
\begin{equation}
\begin{aligned}
    D_{\text{2010}} & \equiv \text{LCSR} \oplus \text{BaBar}_{2010} \oplus \text{Belle}_{2010}\,,
    &               & \text{\cite{delAmoSanchez:2010zd,Ha:2010rf}} \\
    D_{\text{2013}} & \equiv \text{LCSR} \oplus \text{BaBar}_{2012} \oplus \text{Belle}_{2013}\,.
    &               & \text{\cite{Lees:2012vv,Sibidanov:2013rkk}}  \\
\end{aligned}
\end{equation}
Their respective likelihoods are formed as product of $P(\text{LCSR}|\vec x)$ (compare \refeq{LHLCSR})
with the individual experimental likelihoods. We use exclusively the experimental
results on the decay $\bar{B}^0\to \pi^+\ell^-\bar\nu_\ell$, with kinematical cuts $0 \leq q^2 \leq 12\,\GeV^2$.
All experiments \cite{Lees:2012vv,Sibidanov:2013rkk,delAmoSanchez:2010zd,Ha:2010rf} provide their
results as mean values $\vec\mu^E$ for six $q^2$ bins
of width $2\,\GeV^2$, and include sufficient information to construct the covariance matrices $\Sigma^E$.
Thus, we use $P(E|\vec x) = \mathcal{N}_6(\vec\mu^E, \Sigma^E|\vec x)$, for $E = \text{BaBar}_{2010},
\text{Belle}_{2010}, \text{BaBar}_{2012}$ and $\text{Belle}_{2013}$, respectively.

We obtain posterior distributions from our prior and the likelihoods. From the posteriors follow two
best-fit points for the individual experimental data sets,
\begin{equation}
\begin{aligned}
    \vec x^*_\text{2010} & = \argmax P(\vec x|D_{2010}) = (3.44\cdot 10^{-3}, 0.281, -2.14, -0.364)\,,\\
    \vec x^*_\text{2013} & = \argmax P(\vec x|D_{2013}) = (3.33\cdot 10^{-3}, 0.288, -1.94, -0.465)\,,\\
\end{aligned}
\end{equation}
When compared to the best-fit point $\lambda^*_\text{LCSR}$ (see \refeq{BCLbfp}), the above two points exhibit
a marked negative shift in the parameter $b_1^+$ of $-0.89 \simeq 74\%$ and $-0.69 \simeq 58\%$ for the 2010 and 2013 data sets, respectively.
In order to calculate the goodness of fit, we assume $N_\text{d.o.f.} = 14$ degrees of freedom,
which follows from 18 observations (6 theoretical inputs and 12 experimental observations), reduced by
$\dim \vec x = 4$ fit parameters. The p values follow from a $\chi^2$-distribution with
$N_\text{d.o.f.}$ degrees of freedom for the pull values at the respective best-fit point.

\begin{figure*}[t]
\begin{tabular}{cc}
    \subfigure[]{\includegraphics[width=0.48\textwidth]{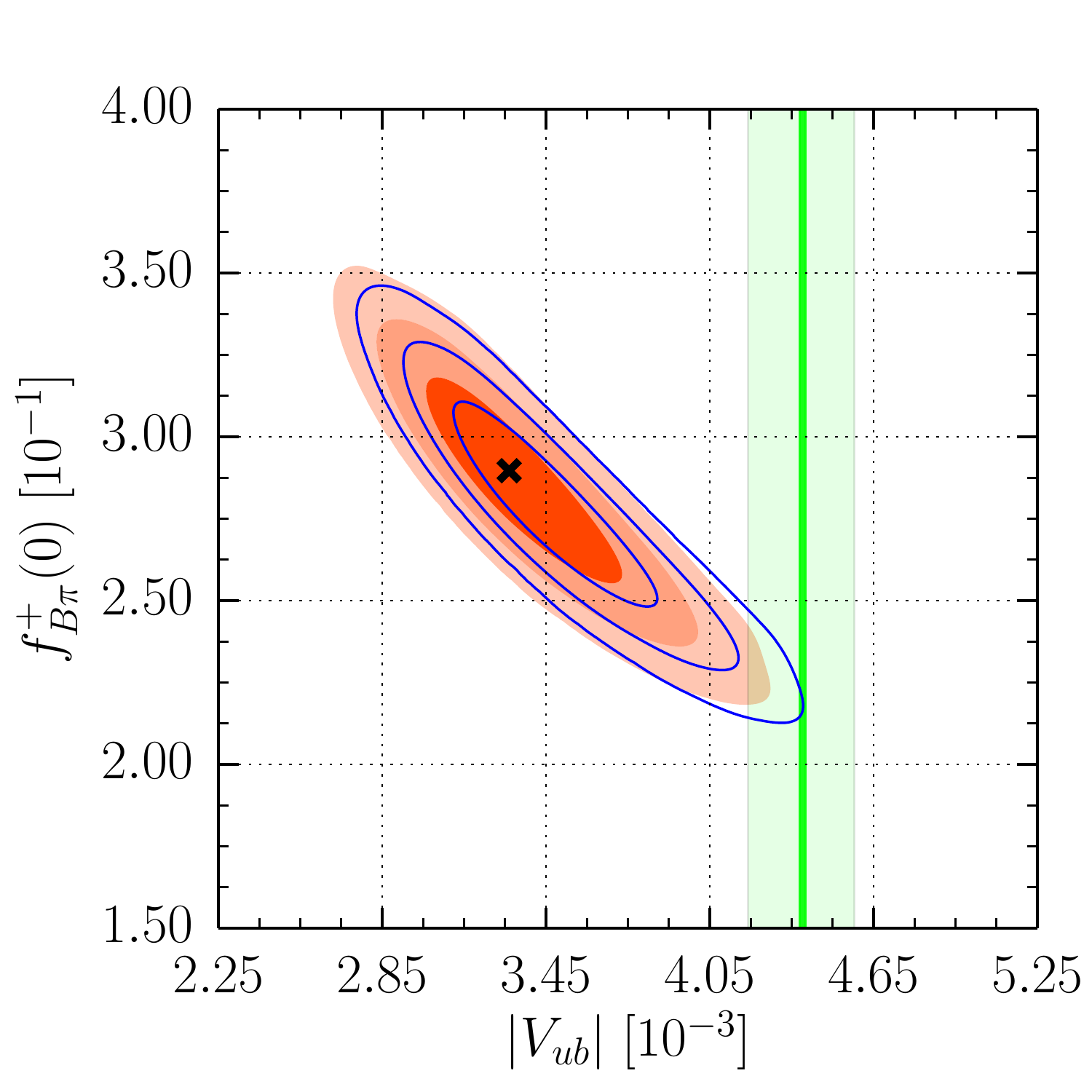}} &
    \subfigure[]{\includegraphics[width=0.48\textwidth]{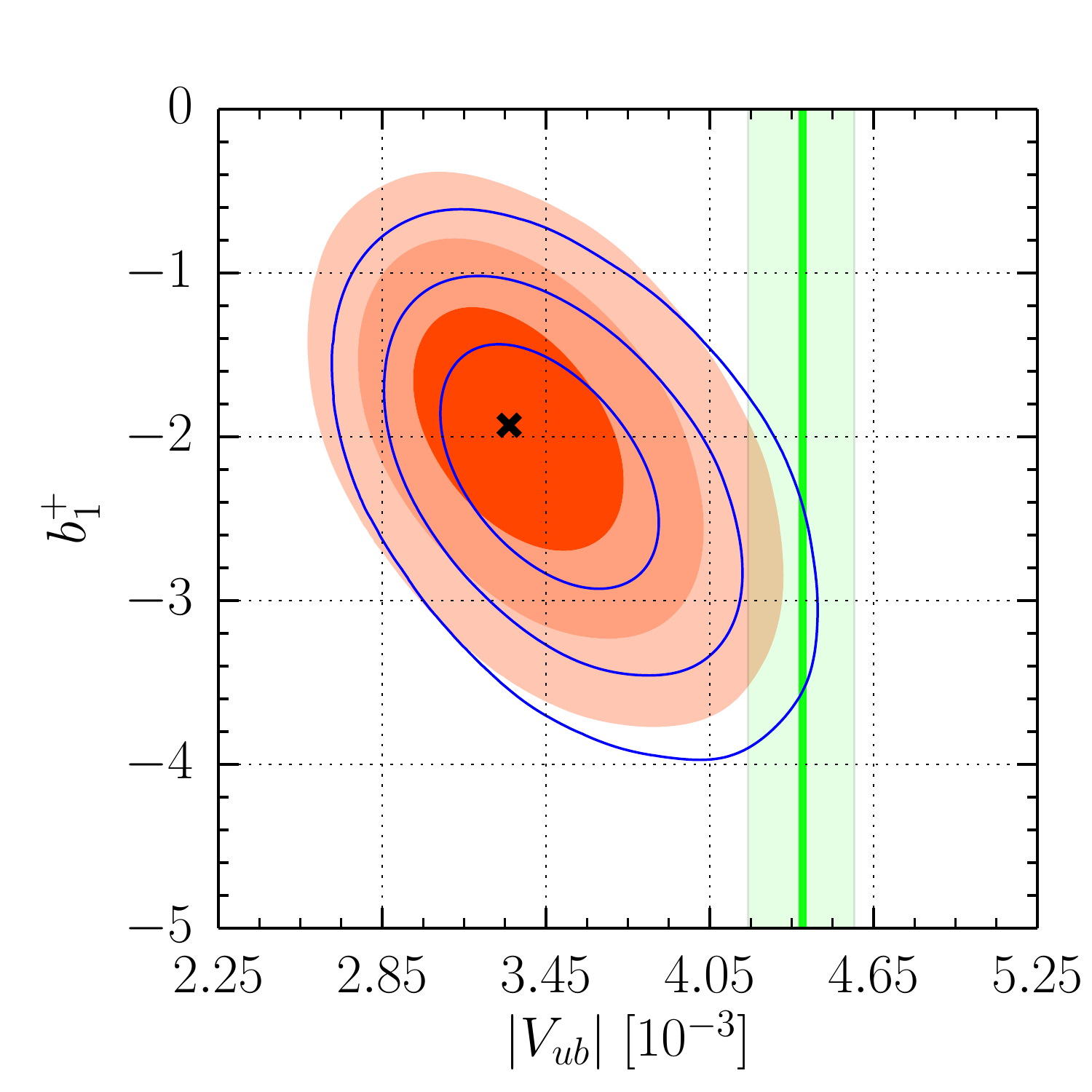}} \\
    \subfigure[]{\includegraphics[width=0.48\textwidth]{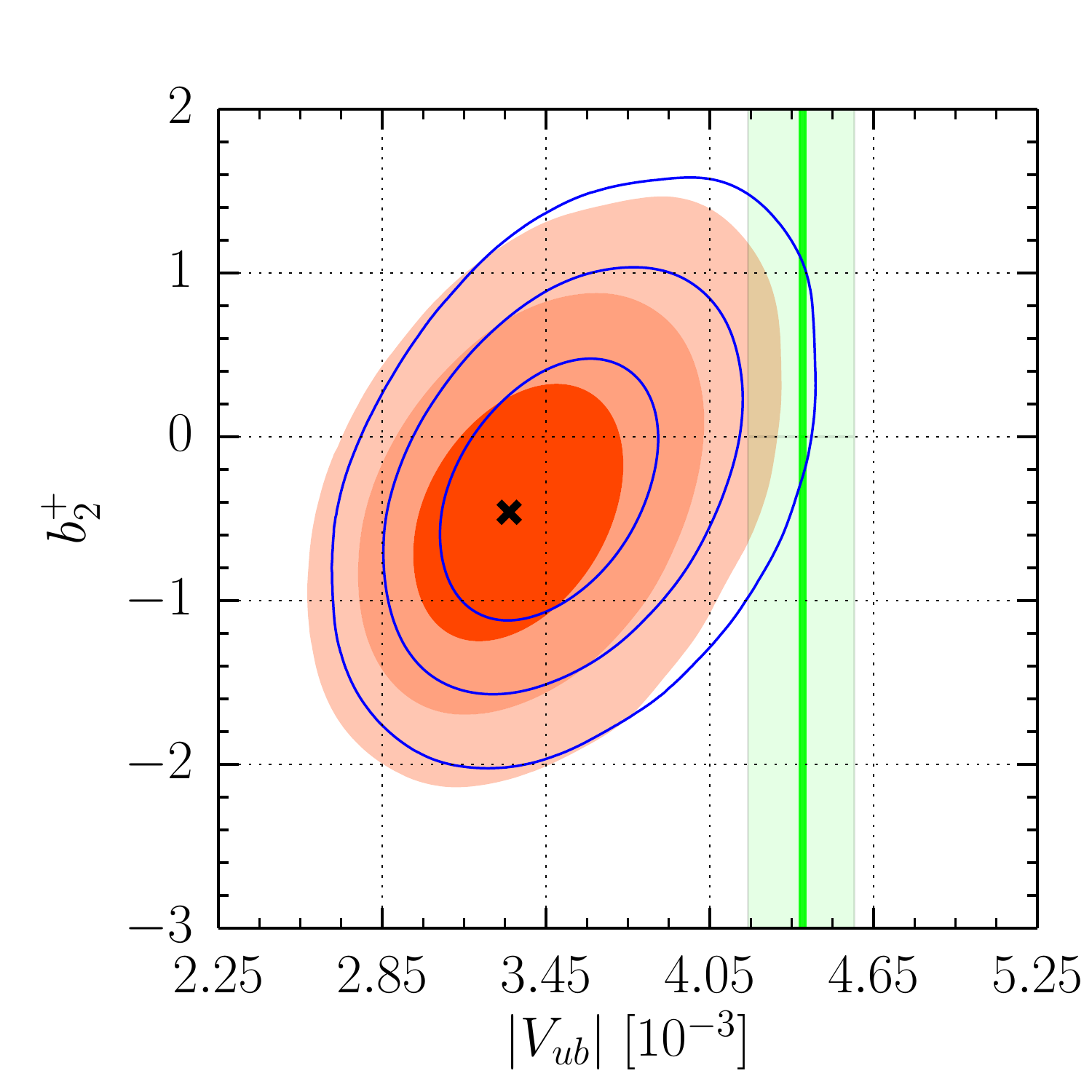}} &
    \begin{minipage}[b][.48\textwidth][t]{.48\textwidth}
    \caption{The two-dimensional marginal posteriors for $|V_{ub}|$ versus the BCL parameters (a) $f_{B\pi}^+(0)$,
    (b) $b_1^+$, and (c) $b_2^+$. The dark orange, orange, and light orange regions show, respectively, the
    $68\%$, $95\%$ and $99\%$ probability regions when using the ``2013'' data set. The
    blue contours delineate the corresponding probability regions of the ``2010'' data set.
    The green and light green vertical bands denote the central value and $68\%$ CL interval of the
    HFAG world average \cite{Amhis:2012bh} of the $|V_{ub}|$ determinations from inclusive
    decays $B\to X_u\ell\bar\nu$ according to the GGOU method \cite{Gambino:2007rp}.
    \label{fig:2Dmarginals}
    }
    \end{minipage}
\end{tabular}
\end{figure*}

We find a good fit, with p values $0.98$ and $0.42$ for the ``2010'' and ``2013'' data sets, respectively. We proceed to calculate the Bayes factor for both data sets, and obtain
\begin{equation}
    \frac{P(D_{2010})}{P(D_{2013})}
    \equiv \frac{\int \rmdx{\vec{x}} P(D_{2010}|\vec{x}) P_0(\vec{x})}{\int \rmdx{\vec{x}} P(D_{2013}|\vec{x}) P_0(\vec{x})} = 3.8\cdot 10^4\,.
\end{equation}
The hypothesis ``LCSR results are in agreement with 2010 data'' is therefore decisively favoured
over the hypothesis ``LCSR results are in agreement with 2013 data''.
The remainder of the goodness-of-fit values for both data sets is displayed in \reftab{gof}.

We show the $68\%$, $95\%$ and $99\%$ probability regions of those two-dimensional marginal distribution
that involve $|V_{ub}|$ for both data sets in \reffig{2Dmarginals}. There we also compare our determination of $|V_{ub}|$
from $B^0\to \pi^+\ell\bar\nu_\ell$ with the HFAG world average of the inclusive determination according
to GGOU \cite{Gambino:2007rp},
\begin{equation}
    |V_{ub}|^\text{HFAG,GGOU} = (4.39 \pm 0.15 {}^{+0.12}_{-0.14})\cdot 10^{-3}\,.
\end{equation}
We find that the ``2010'' data set is compatible with the inclusive determination at the 3$\sigma$
level. However, the ``2013'' data set moves even further away from the inclusive values.

Through integration over all BCL parameters we obtain the one-dimensional marginal posteriors $P(|V_{ub}|\,|D_{2010})$
and $P(|V_{ub}|\,|D_{2013})$.\footnote{%
Variates of either posterior distribution can be obtained from the authors upon request.
}
Their respective $68\%$ probability intervals read
\begin{equation}
\begin{aligned}
    |V_{ub}|^\text{2010} & = (3.43^{+0.27}_{-0.23}) \cdot 10^{-3}\,,\\
    |V_{ub}|^\text{2013} & = (3.32^{+0.26}_{-0.22}) \cdot 10^{-3}\,.\\
\end{aligned}
\end{equation}
Both intervals are compatible with each other at a level of less than 1$\sigma$.

\section{Conclusion \label{sec:concl}}

We have carried out the first Bayesian analysis of the
$B\to \pi$ vector form factor within the framework of LCSR in QCD.
For this, it was instrumental to construct a likelihood that
relates the sum rule to the experimentally measured $B$-meson
mass. As a consequence, our analysis yields correlated constraints on the input parameter space.
One of our main results are predictions for the form factor $f_{B\pi}^+(q^2) $ and its
derivatives at two separate values of momentum transfer $q^2 = 0$ and $10\,\GeV^2$,
well within the window of applicability of the LCSR.
A comprehensive ''diagnostics''
of the obtained probability distributions for each input parameter is described above.

Based on these results,
we obtain a joint posterior-predictive probability distribution for the
parameters of the $z$-series representation.
This distribution is urgently needed for the precise extrapolation of
$f_{B\pi}^+(q^2)$ toward large $q^2$, beyond the LCSR region.
Interestingly, we find theoretical uncertainties that are about 20\% smaller than those
obtained in previous analyses, where only naive estimates --- based on individual
variations of each input parameter --- had been carried out. Especially encouraging is a
reasonably small dependence on the combined renormalization/factorization scale, which
was separately investigated by varying the combined scale in the same interval
as in \cite{Khodjamirian:2011ub}.
All these findings prompt the conclusion that a simultaneous ``scanning'' of the
input parameter space within a Bayesian analysis is probably the only revealing way to assess
the realistic uncertainties of a non-lattice QCD-method such as LCSR.
Our analysis supports the use of the $B$-meson mass to constrain
the effective threshold interval in the quark-hadron duality approximation.

A word of caution should be added to the above comments, reminding 
that the accuracy estimated in this paper  concerns a certain {\it approximation} 
of LCSR, with a truncated OPE and the quark-hadron duality ansatz applied 
to the correlation function and to the rigorous hadronic dispersion relation.
A further improvement  of LCSRs is desirable, but demands several technically 
challenging computations:
the complete NNLO corrections to twist 2 and 3 terms; the nonasymptotic corrections
to the twist-3 NLO part; and an assessment of twist-5 and 6 terms. On the side of the
input parameters, it is desirable to improve our knowledge of
the pion DAs. Important constraints on their parameters arise from other LCSRs,
such as the ones for the pion electromagnetic form factor and the 
$\gamma \gamma^* \to \pi$ form factor.
In fact, even more information on the correlations between the various input parameters
could be obtained from a global analysis; e.g., by incorporating measurements
of the aforementioned form factors into the likelihood.

Turning to the comparison with other theoretical predictions of the $B\to \pi$
form factor, we notice that our results at low $q^2$  are consistent with the 
outcome of the previous analyses of LCSR, if one adopts the 
simplified uncertainty estimates in these analyses.
Furthermore, our extrapolation to large $q^2$ is in a very good agreement with
the published lattice QCD results.

As a byproduct of our analysis, we can predict the strong $B^*B\pi$
coupling. Our result is in the ballpark of earlier direct LCSR calculations
based on double dispersion relations with simple duality ansatz  and obtained from a 
less accurate correlation function. It is therefore important to
update the latter calculation and include it in one statistical pool 
with the $B\to \pi$ form factor.

The second main result of this paper is the implementation of the model-independent bounds
for the form factor, which allow one to confirm the reliability of the extrapolation that is
based on truncated $z$-series. We studied different versions of these bounds and found that 
the ones which include form factor and its first and second derivative (all at one value of $q^2$)
are the most confining and useful ones. We obtain an upper/lower bound at $q^2=20\,\GeV^2$
that is only about $\pm 25\%$ larger/smaller than the average value of the extrapolated form factor.
Our findings will be important for the comparison with the respective lattice QCD results,
which can currently be calculated at $q^2 \gtrsim 17\,\GeV^2$.

The third main result of this paper is the determination of $|V_{ub}|$ from available
experimental data within the region of LCSR. 
We find that the two sets of experimental
analyses are very compatible with the LCSR predictions for the vector form factor. However,
based on a Bayesian model comparison, we also find that 2010 data set is in decisively better
agreement with the theory predictions than the 2013 data set.

Note that given the approximation of LCSR, the theoretical uncertainty in $|V_{ub}|$  
obtained form our analysis is comparable to the one from the most accurate 
determinations of this CKM parameter in the inclusive $b\to u$ transitions.
We find that our results exhibit a tension
with respect to the GGOU determination beyond the level of $99\%$ probability.

Concluding, we foresee an immediate extension 
of this work to other exclusive $b\to s $ and $b\to u$ 
transitions, comprehensively updating the LCSRs 
for the $B\to K$ and $B_s\to K$ form factors and applying the statistical analysis.

\acknowledgments

D.v.D.\ would like to thank Frederik Beaujean for helpful discussions.
I.S.I.\ would like to thank Irinel Caprini for useful discussions.
This work is supported in parts by the Bundesministerium f\"ur Bildung und Forschung
(BMBF), and by the Deutsche Forschungsgemeinschaft (DFG), Research Unit FOR
1873 (“Quark Flavour Physics and Effective Field Theories”),
Contract No.KH 205/2-1.

\appendix

\section{Formulae relevant for the unitarity bounds}

\paragraph{1.} The expression for the outer function
at $\tilde{q}^2\neq 0$ and $n+1$ differentiations is
obtained after transformation of the variable $t\to z(t,t_0) $ in the kinematical factor
 $k_T(t,\tilde{q}^2,n)$ entering the integral (\ref{eq:fBpicontrib}) including
also the Jacobian of this transformation:
\begin{equation}
\label{eq:wrho}
|\phi(z,\tilde{q}^2,n)|^2 \sim  k_T(t(z),\tilde{q}^2,n)\,
\frac{dt(z,t_0)}{dz}\,,
\end{equation}
To obtain a function $\phi(z,\tilde{q}^2,n)$ with desired analytical properties
(no poles and/or zeros inside the unit disc) one has to multiply the above expression 
by  unimodular functions that are equal 1
on the unit circle, hence do not change the value of the integral. 
We skip this part of the derivation for brevity. In addition,
in order to eliminate the $B^*$ pole located on the real axis of $z$ plane,
the outer function is also multiplied by the (unimodular) Blaschke factor 
\cite{Lellouch:1995yv,Bourrely:2008za},
\begin{eqnarray}
B(z,t_0) = \frac{z-z(m_{B^*}^2,t_0)}{1-z\,z(m_{B^*}^2,t_0)}\,.
\end{eqnarray}
As a result, the outer function at $\tilde{q}^2 \neq 0$ and general number of differentiations 
$n$ is:
\begin{eqnarray}
\label{eqn:OF1}
\phi(z,\tilde{q}^2,n) = 
\frac{B(z,t_0)}{\sqrt{32\pi}}\Big(\sqrt{t_+-t(z,t_0)}+ \sqrt{t_+-t_0}\Big)\nonumber\\
\times \frac{(t_+-t(z,t_0))}{(t_+-t_0)^{1/4}}
\frac{(\sqrt{t_+-t_-}+ \sqrt{t_+-t(z,t_0)})^{3/2}}{
(\sqrt{t_+}+ \sqrt{t_+-t(z,t_0)})^2
(\sqrt{t_+ +\tilde{q}^2}+ \sqrt{t_+-t(z,t_0)})^{n+2}
}
\end{eqnarray}

At $n=1$ and $\tilde{q}^2=0$ it coinsides with the expression given in \cite{Bourrely:2008za}.\\[2mm]

\paragraph{2.} The $5\times5$ matrix used for the derivation of the bounds
as decribed in the text  has the following expression
\be
{\cal M}= \left(
 \begin{array}{c c c c c c}
 \langle h|h\rangle			&h_0 				&h_0+h_1\epsilon+h_2\epsilon^2	 	&h_0-h_1\epsilon+h_2\epsilon^2			& h(z)\\
 h_0 					&\frac{1}{1-z_0^2}		&\frac{1}{1-z_0 (z_0+\epsilon )}		&\frac{1}{1-z_0 (z_0-\epsilon )}			&\frac{1}{1-z z_0}	\\
 h_0+h_1\epsilon+h_2\epsilon^2 		&\frac{1}{1-z_0 (z_0+\epsilon )} 	&\frac{1}{1-(z_0+\epsilon)^2}		&\frac{1}{1-(z_0+\epsilon )(z_0-\epsilon )}	&\frac{1}{1-z ( z_0+\epsilon)} \\
 h_0-h_1\epsilon+h_2\epsilon^2		&\frac{1}{1-z_0 (z_0-\epsilon )}	&\frac{1}{1-(z_0+\epsilon )(z_0-\epsilon )}	&\frac{1}{1-(z_0-\epsilon)^2}		&\frac{1}{1-z ( z_0-\epsilon)} \\
 h(z)  		 			&\frac{1}{1-z z_0}		&\frac{1}{1-z (z_0+\epsilon)}		&\frac{1}{1-z (z_0-\epsilon)}			&\frac{1}{1-z^2}\\
 	\end{array}\right)
 \ee
 where $h_{0,1,2}$ are the shorthand notations introduced in \refeq{h012}.

\bibliographystyle{JHEP}
\bibliography{references.bib}

\end{document}